\documentclass[amsmath,amssymb,twocolumn,superscriptaddress]{revtex4}
\usepackage{graphicx}% Include figure files
\usepackage{dcolumn}% Align table columns on decimal point

\def\be{\begin{equation}}
\def\ee{\end{equation}}
\def\beq{\begin{eqnarray}}
\def\eeq{\end{eqnarray}}

\usepackage{bm}% bold math
\usepackage{graphicx}
\usepackage{multirow}
\usepackage{dcolumn}
\usepackage{amsmath}
\usepackage[latin1]{inputenc}
\usepackage{graphicx, psfrag}
\usepackage{amssymb}
\usepackage[colorlinks=true, citecolor=blue, urlcolor = blue, linkcolor= red, bookmarks=true]{hyperref}
\usepackage{float}
\usepackage{amsmath}
\usepackage{amsfonts}
\usepackage{dcolumn}
\usepackage{hyperref}
\usepackage{subfigure}
\usepackage{pgfplots}
\usepackage{epstopdf}
\usepackage{booktabs}

\begin{document}

%\preprint{}

\title{Exploring physical properties of minimally deformed strange star model and constraints on maximum mass limit in $f(\mathcal{Q})$ gravity} 



\author{S. K. Maurya}%
\email[Email:]{sunil@unizwa.edu.om}
\affiliation{Department of Mathematical and Physical Sciences,
College of Arts and Sciences, University of Nizwa, Nizwa, Sultanate of Oman}

\author{G. Mustafa} \email[Email:]{gmustafa3828@gmail.com} 
\affiliation{Department of Mathematics, Shanghai University,
Shanghai, 200444, Shanghai, People's Republic of China}

\author{M. Govender}
\email[Email:]{megandhreng@dut.ac.za}
\affiliation{Department of Mathematics, Durban University of Technology, Durban 4000, South Africa}

\author{Ksh. Newton Singh} 
\email[Email:]{ntnphy@gmail.com} 
\affiliation{Department of Physics, National Defence Academy, Khadakwasla, Pune 411023, India}

%\thanks{}

\date{\today}

\begin{abstract}
In this work we take our cue from the observations of gravitational waves of the GW190814 event which suggests that source of the signals can be ascribed to a compact binary coalescence of a 22.2 to 24.3$ M_{\odot}$ black hole and a compact object endowed with a mass of 2.50 to 2.67$M_{\odot}$. In the current exposition, we are concerned with modeling of the lower mass component of the coalescence pair. We utilise the  $f(\mathcal{Q})$ gravity together with Minimum Geometric Deformation (MGD) technique to obtain compact stellar objects with masses aligned with the GW190814 event. Starting off with the Tolman IV ansatz for one of the metric functions, together with a MIT Bag model equation of state we are able to reduce the problem of fully describing the gravitational behaviour of the seed solution to a quadrature. Through the MGD technique, we introduce anisotropy by deforming the radial part of the gravitational potential. This enables us to obtain two new classes of solutions which depend on the metricity parameter, $\cal Q$ and the deformation constant, $\beta$. We show that these two parameters play a crucial role in determining the thermodynamical behaviour and stability of our models. In particular, we show that the interplay between the metricity parameter and the deformation constant leads to predicted mass of the progenitor articulating as the secondary component of  GW190814.
\end{abstract}

%\keywords{}

\maketitle


\section{Introduction}
Modified theories of gravity continue to attract widespread attention amongst astrophysicists and cosmologists alike. While Einstein's gravitational theory has rewarded us handsomely with explanations of physical observations such as the deflection of starlight in the presence of a massive gravitating body or the accidental discovery of the Cosmic Microwave Background Radiation, it falls short in accounting for various other physical phenomena\citep{misner,shapiro,ellis}. This has prompted researchers to adopt alternative theories of gravity or modifications to classical general relativity. Observations of peculiar stellar characteristics such as high redshifts or mass-radius relations beyond the theoretical bounds conjured up the need for exotic matter fields beyond the standard model. Apart from pressure anisotropy, density inhomogeneities, electric charge, bulk and shear viscosities, dark energy, dark matter, strings and various scalar fields were thrown into the mix to account for observational data \cite{dirac1,dirac2,brans1,fa04}. The foregoing decade has witnessed a massive influx of both cosmological and astrophysical models borne out of modified gravity theories. These offsprings of Einstein's general relativity have continued to enjoy grandiose success on the theoretical front in terms of offering alternative explanations for both cosmological as well as astrophysical observations. 

 One of the earlier alternatives to classical general relativity was a scalar-tensor theory of gravity, the so-called Brans-Dicke theory \cite{brans1}. The non-minimally coupled scalar field masquerades as the spacetime-varying gravitational "constant" and has been successfully utilised in explaining inflation \cite{guth} (without invoking the use of magnetic monopoles or the inflaton) and observed late-time acceleration of the Universe.  Any extension or modification to classical general relativity must have as its limit Einsteinian gravity. This requirement serves as a springboard for modified gravity theories which include $f(R)$, $f(R,T)$, $f(\mathcal{Q})$ and Lovelock gravity amongst others can be found in Refs. \cite{harko,hans-ban,Rastall,Rastall1,unimod1,unimod2,unimod3,ellis1}.
 
The $f(T)$ gravity theory arises from Einstein's construction of the {\em Teleparallel Equivalent of General Relativity} (TEGR) where gravity arises from torsion. The torsion scalar, $T$ is obtained by contracting the torsion tensor, which when used as a Lagrangian produces the field equations of general relativity. Naturally, if one extends $T$ to $f(T)$ in the Langranian one obtains $f(T)$ modified gravity which arises from TEGR \cite{r5,r6}. Cosmologists have derived a fair amount of success explaining the inflationary epoch and late-time acceleration of the Universe within the $f(T)$ framework. In the so-called symmetric teleparallel general relativity (STGR) \cite{r7,r8,r9} which is an equivalent description of general relativity as TEGR, both the curvature and torsion vanish. In this formalism it is the nonmetricity $\mathcal{Q}$ which describes gravity. Furthermore, as in the $f(T)$ framework \cite{r14,r15,r16}, the gauge choice leads to the loss of the coordinate transformation invariant in $f(\mathcal{Q})$ gravity, giving rise to different consequences which depend on the coordinate system employed\cite{Zhao}. \\
 
A natural extension of STGR is the $f(\mathcal{Q})$ gravity which provides us with a simpler geometrical formulation
of classical GR in which the affine spacetime structure has no bearing, thus incising the inertial character of the gravitational interaction. Recently,  $f(\mathcal{Q})$ gravity has attracted the attention of cosmologists and astrophysicists, with more applications in cosmology and astrophysics coming to the fore. It has been demonstrated that the  $f(\mathcal{Q})$ gravity can account for analytical cosmological solutions describing the acceleration of the Universe for early and late epochs of its evolution without resorting to exotic matter content such as scalar fields \cite{Koivisto,sahoo1,sahoo2}. Models of holographic dark energy in the framework of $f(\mathcal{Q})$ gravity have been explored\cite{Shekh1}. By adopting a functional form $f(\mathcal{Q}) = \mathcal{Q} + \mathcal{Q}^n$, together with a relation between the cosmic time and cosmological redshift, several models of inflation driven by dark energy were presented. It was shown that these models can account for the recently observed state-finder parameters. In a recent exposition, static spherically symmetric solutions in $f(\mathcal{Q})$ gravity have been obtained. Assuming a constant metricity scalar, ${\cal{Q}} = {\cal Q}_0$, several exact solutions of the governing equations for an anisotropic fluid are obtained. It was further demonstrated that there does not exist a  Schwarzschild analogue solution for nonvanishing $f(\mathcal{Q})$ function\cite{Wang}. The $f(\mathcal{Q})$ formalism has been successfully utilised to derive wormhole solutions\cite{r23}, explore modified 
energy conditions \cite{r24} and to investigate the consequences of the theory in the Post-Newtonian limit \cite{r25}. 
 
In recent investigations, the concept of gravitational decoupling (GD) has been incorporated into the Einstein-Gauss-Bonnet (EGB) framework. As in the standard 4D classical gravity theory, GD allows for anisotropisation of seed solutions thus providing a mechanism to study the impact of anisotropic stresses in compact objects. The minimal geometric deformation (MGD) method \cite{o1} was utilised to model a compact star in 5D EGB gravity. This work showed that synergistic contributions from the decoupling parameter and the EGB constant lead to higher neutron stars masses \cite{sunil5d4}. In this connection, the extended MGD methodology \cite{o2ol} was applied to investigate the exact solution for compact star in the framework of 5D EGB gravity by Maurya et al.\cite{MGD29}. \\
There has been a plethora of research projects on modeling of compact objects such as neutron stars and strange stars employing MGD and complete geometric deformation (CGD). The reader is referred to these references and works cited within for an up-to-date view of the applications of gravitational decoupling in an astrophysical setting in the following works \cite{MGD1,MGD7,MGD8,MGD9,MGD13,MGD27,MGD2,MGD3,MGD4,MGD12,MGD5, MGD6}. In this connection some pioneering works on gravitational decoupling can been seen in Refs. \cite{MGD10,MGD26,MGD14,MGD15,MGD16,MGD17,MGD18,MGD19,MGD20,MGD21,MGD22,MGD23,MGD25,MGD24,MGD251,MGD30,MGD28,MGD31}.  

There have been numerous theoretical models of compact objects such as neutron stars that have been presented in the literature offering possible explanations for the gravitational wave detection events, both in classical general relativity as well as modified gravity theories. The first detection of gravitational waves in August of 2017 arising from the merger of two neutron stars (GW170817 event) has fuelled speculation as to the nature of the sources giving rise to the observed signals. The GW170817 event was thought to be the result of a binary neutron star (NS) inspiral with masses in the range of 1.17 - 1.60$ M_{\odot}$. Furthermore, the source of GW170817 also produced two strong electromagnetic signals, the first being a short gamma-ray burst GRB170817A with a ~2 s delay with respect to the gravitational wave signal and a kilonova, catalogued as AT2017gfo, with the intensity of its luminosity peaking a few days after the merger \cite{abbotta,abbottb,abbottc}. The electromagnetic footprint of GW170817 ruled out very soft or very stiff equations of state \cite{burgio}. 

Researchers have appealed to modified theories of gravity to construct models of compact objects with mass and radii limits beyond the realm of current observations in the hope of explaining the various gravitational wave detection events. In this vein, recent efforts by \cite{r99,r100} have proved fruitful in obtaining noteworthy anisotropic solutions for compact stars and charged spherically symmetric black holes in the framework of the $f(R)$ gravity along with their stability analysis. In this connection, Astashenok and his collaborators \cite{Astashenok2020,Astashenok2021} show that the NS of mass 2.67$M_{\odot}$ can be described with the mass-radius relation obtained by Extended Theories of Gravity. They also argued that masses of rotating neutron stars can exceed 2.67$M_{\odot}$ for some equations of state. Furthermore, Maurya et al. \cite{MGD23} have demonstrated the existence of strange star candidates beyond the standard maximum mass limit by employing the gravitational decoupling method within 5D Einstein-Gauss-Bonnet formalism.

A peculiar gravitational wave detection made by the LIGO-Virgo collaboration referred to as the GW190814 event suggests that the signals arising from this observation were due to a compact binary coalescence of a $23.2^{+1.1}_{-1.0}{M_\odot}$ black hole (BH) and a compact object endowed with a mass of 2.50 to 2.67$M_{\odot}$\cite{wenbin}. The source giving rise to these gravitational waves has the most unequal mass ratio to date, $0.112_{-0.009}^{+0.008}$. The real mystery that arises from this event  points to the secondary component as either being the lightest black hole or the heaviest NS star ever observed in a binary compact-object system. 
While the primary component of GW190814 is widely accepted to be a BH, it is the nature of the secondary component that has spurned some deeper diving into limits on mass-radii relations of compact objects. It has been pointed out that the lack of measurable tidal deformations and the absence of an electromagnetic component, points to the low-mass member being either an NS or a BH. 
However, comparison with the nature of the secondary component with the GW170817 event indicates that the secondary in GW190814 is too heavy to qualify as an NS. Researchers are left with the antithesis that the secondary is either the lightest BH or the heaviest NS ever observed in a binary system.
Within the context of classical GR it is still a challenge to account for a secondary component with a 2.67$M_{\odot}$ neutron star in GW190814 despite what researchers have predicted from the binary neutron star (bNS) merger event GW170817. An interesting approach to produce the low-mass component of GW190814 is to view it as a bNS merger remnant. This secondary merger can push up the mass limit of the low-mass component to approximately 3.4$M_{\odot}$. These 
2nd-generation mergers have been widely explored in several contexts including scattering of a NS-NS tightly coupled system scattering of a massive BH or a hierarchical triple system comprising of an initial bNS coalescence, the remnant of which merges with a 23 $M_{\odot}$ black hole (see \cite{wenbin} and references therein.).

Our current work attempts to describe the smaller body of the binary by utilising the $f(\mathcal{Q})$ gravitational theory.  We employ the $f(\mathcal{Q})$ framework together with the MGD approach to model the secondary compact object which was part of the source of gravitational waves observed by the LIGO-Virgo group. In order to close the system of equations describing the seed solution, we employ the Tolman IV ansatz for one of the metric potentials in addition to utilising the MIT Bag model equation of state. Both the seed solution and the solution arising from the extra source term due to MGD are anisotropic. We investigate the role played by anisotropy in the observed radius and mass of the secondary component of event GW190814. 

The article is organized as follows:  Section \ref{sec2} consists of a detailed review of the field equations in $f(\mathcal{Q})$ gravity theory together with gravitational decoupling methodology via MGD approach by introducing an extra source. The energy-momentum tensor under different sources with the MIT Bag equation of state (EoS) have been also discussed. In Sec. \ref{sec3} we obtain a minimally deformed  solution by employing a well-behaved Tolman IV metric potential for the seed spacetime geometry to ensure a well-defined horizon-free spacetime.  The key task here is to determine classes of non-singular solutions with a well-behaved deformation function $\psi(r)$. To achieve this, two different procedures have been proposed in subsections \ref{sec3.1} and \ref{sec3.2}, namely via the mimic constraint approach, to close the system of equations for the extra source introduced by gravitational decoupling. The exterior spacetime and junction conditions have been discussed in Sec.\ref{sec4} in which we join the minimally deformed anisotropic interior solution in $f(\mathcal{Q})$ gravity to the exterior Schwarzschild Anti-de Sitter vacuum  solution at a pressure-free boundary. The physical analysis of the minimally deformed solution for strange star (SS) obtained in subsections \ref{sec3.1} and \ref{sec3.2} have been discussed in  different subsections under Sec.\ref{sec5}. The regularity condition of the SS model is discussed in subsection \ref{sec5.1} while its stability analysis is done in subsections \ref{sec5.2} and \ref{sec5.3}. The most important physical parameters such as the measurements of the mass ($M$) and radius $(R$) of the SS model have been determined via the $M-R$ curves in Sec. \ref{sec5.4}, while the constraints on the maximum mass limit and Bag constant of the SS models via equi-plane diagrams can be seen in Sec. \ref{sec5.5}. In the last Sec. \ref{sec6}, we present a discussion of findings together with some astrophysical implications of the models.  Finally, some relevant lengthy expressions of physical quantities are presented in the Appendix.

 \section{Field equations for $f(\mathcal{Q})$ gravity with extra source}\label{sec2}
 
According to the standard description of general relativity, the Levi-Civita affine connection on the spacetime manifold is metric compatible. On any manifold, however, alternative affine connections may provide various insight features and these connections can result in distinct but similar explanations of gravity \cite{rm1,rm2}. Except for curvature $R$, the Levi-Civita connection adopted by GR requires that the other two essential geometrical concepts, nonmetricity $\mathcal{Q}$ and torsion $T$, both vanish. By relaxing these requirements, it is theoretically possible to build theories of gravity based on non-Riemannian geometry with nonvanishing curvature, torsion and nonmetricity. One can define the teleparallel gravity theory equivalent of GR by choosing a connection that requires both curvature and nonmetricity to disappear while relaxing the torsion restriction \cite{r6}. The major difference among symmetric teleparallel gravity and GR is the affine connection. A third option is to examine a spacetime manifold without torsion but with nonvanishing nonmetricity, which leads to the symmetric teleparallel formulation of GR \cite{rm3,rm4,rm5,r6}. The $f(\mathcal{Q})$ gravity may be formulated by considering a gravitational Lagrangian that contains an arbitrary function of the nonmetricity $\mathcal{Q}$. The expanding history of the Universe in $f(\mathcal{Q})$ gravity is one of the key motives for this extension. Herein, we are going to present a detailed discussion about the modified $f(\mathcal{Q})$ gravity for gravitationally decoupled system. The symmetric teleparallel theory, i.e., $f(\mathcal{Q})$ gravity was originally given by Jimenez et al. \cite{r1}. In $f(\mathcal{Q})$ gravity, the nonmetricity scalar $\mathcal{Q}$ drives the gravitational interaction. The action for modified $f(\mathcal{Q})$ gravity for a  gravitationally decoupled system can be expressed by including an extra Lagrangian $L_\theta$ for another source $\theta_{\epsilon \varepsilon}$ as:

\begin{small}
\begin{eqnarray}
\label{eq1}
\mathcal{S}=\underbrace{\int\frac{1}{2}\,f(\mathcal{Q})\sqrt{-g}\,d^4x+\int \mathcal{L}_m\,\sqrt{-g}\,d^4x}_{\mathcal{S}_{Q}}+\beta\underbrace{ \int \mathcal{L}_\theta\,\sqrt{-g}\,d^4x}_{\mathcal{S}_{\theta}},~~~
\end{eqnarray}
\end{small}
where $\mathcal{L}_m$ represents the Lagrangian density of matter fields appearing in the  $f(\mathcal{Q})$ gravity theory corresponding to energy momentum tensor $T_{\epsilon\varepsilon}$ and $\mathcal{L}_{\theta}$ denotes a Lagrangian density of a new gravitational sector which is not described by  $f(\mathcal{Q})$ gravity, let us call a  "$\theta$-gravitational sector" ($\theta_{\epsilon\varepsilon}$). This new extra contribution can always introduce a corrections to matter fields of the symmetric teleparallel gravity and it can be consolidated as part of an effective energy-momentum tensor $T^{\text{eff}}_{\epsilon\varepsilon}=\big(T_{\epsilon\varepsilon}+\beta\,\theta_{\epsilon\varepsilon}\big)$. Furthermore, $g$ denotes the determinant of the metric tensor $g_{\epsilon\varepsilon}$, and $\beta$ is a decoupling constant. The  nonmetricity tensor $\mathcal{Q}$ is defined by the following relation
\begin{equation}\label{eq2}
\mathcal{Q}_{\lambda\epsilon\varepsilon}=\bigtriangledown_{\lambda} g_{\epsilon\varepsilon}=\partial_\lambda g_{\epsilon\varepsilon}-\Gamma^\delta_{\,\,\,\lambda \epsilon}g_{\delta \varepsilon}-\Gamma^\delta_{\,\,\,\lambda \varepsilon}g_{\epsilon \delta},
\end{equation}
where $\Gamma^\delta_{\,\,\,\epsilon\varepsilon}$ is known as the affine connection which assumes the form 
\begin{equation}\label{eq3}
\Gamma^\delta_{\,\,\,\epsilon\varepsilon}=\lbrace^\delta_{\,\,\,\epsilon\varepsilon} \rbrace+K^\delta_{\,\,\,\epsilon\varepsilon}+ L^\delta_{\,\,\,\epsilon\varepsilon},
\end{equation}
where $\lbrace^\delta_{\,\,\,\epsilon\varepsilon} \rbrace$, $L^\delta_{\,\,\,\epsilon\varepsilon}$, and $K^\delta_{\,\,\,\epsilon\varepsilon}$ are the Levi-Civita connection, disformation, and contortion tensors respectively, which are determined as:
\begin{eqnarray}
\label{eq4}
&&\hspace{-0.4cm} \lbrace^\delta_{\,\,\,\epsilon\varepsilon} \rbrace=\frac{1}{2}g^{\delta\sigma}\left(\partial_\epsilon g_{\sigma\varepsilon}+\partial_\varepsilon g_{\sigma\epsilon}-\partial_\sigma g_{\epsilon\varepsilon}\right),\nonumber\\
&&\hspace{-0.4cm}
L^\delta_{\,\,\,\epsilon\varepsilon}=\frac{1}{2}\mathcal{Q}^\delta_{\,\,\,\epsilon\varepsilon}-\mathcal{Q}_{(\epsilon\,\,\,\,\,\,\varepsilon)}^{\,\,\,\,\,\,\delta},\nonumber \\
&&\hspace{-0.5cm}  K^\delta_{\,\,\,\epsilon\varepsilon}=\frac{1}{2} T^\delta_{\,\,\,\epsilon\varepsilon}+T_{(\epsilon\,\,\,\,\,\,\varepsilon)}^{\,\,\,\,\,\,\delta},
\end{eqnarray}
with the torsion tensor $T^\delta_{\,\,\,\epsilon\varepsilon}$, which defines the anti-symmetric part of the affine connection, $T^\delta_{\,\,\,\epsilon\varepsilon}=2\Gamma^\lambda_{\,\,\,[\epsilon\varepsilon]}$. The superpotential related to the nonmetricity tensor is defined as:
\begin{equation}\label{eq5}
P^\alpha_{\,\,\,\,\epsilon\varepsilon}=\frac{1}{4}\left[-\mathcal{Q}^\alpha_{\,\,\,\,\epsilon\varepsilon}+2 \mathcal{Q}_{(\epsilon\,\varepsilon)}^\alpha+\mathcal{Q}^\alpha g_{\epsilon\varepsilon}-\tilde{\mathcal{Q}}^\alpha g_{\epsilon\varepsilon}-\delta^\alpha_{(\epsilon}\mathcal{Q}_{\varepsilon)}\right],
\end{equation}
where
\begin{equation}\label{eq6}
\mathcal{Q}_{\alpha}\equiv \mathcal{Q}_{\alpha\,\,\,\epsilon}^{\,\,\epsilon},\; \tilde{\mathcal{Q}}_\alpha=\mathcal{Q}^\epsilon_{\,\,\,\,\alpha\epsilon}.
\end{equation}
are two independence traces, which help us to define the nonmetricity scalar term as
\begin{equation}\label{eq7}
\mathcal{Q}=-\mathcal{Q}_{\alpha\epsilon\varepsilon}\,P^{\alpha\epsilon\varepsilon}.
\end{equation}
In order to derive the field equations for $f(\mathcal{Q})$ gravity, we can set the action Eq. (\ref{eq1}) is constant with respect to the  variation over the metric tensor $g_{\epsilon\varepsilon}$, resulting in
\begin{eqnarray}
\label{eq8}
&& \hspace{-0.5cm}\frac{2}{\sqrt{-g}}\bigtriangledown_\gamma\left(\sqrt{-g}\,f_\mathcal{Q}\,P^\gamma_{\,\,\,\,\epsilon\varepsilon}\right)+\frac{1}{2}g_{\epsilon\varepsilon}f 
+f_\mathcal{Q}\big(P_{\epsilon\gamma i}\,\mathcal{Q}_\varepsilon^{\,\,\,\gamma i}\nonumber\\&&\hspace{-0.4cm}-2\,\mathcal{Q}_{\gamma i \epsilon}\,P^{\gamma i}_{\,\,\,\varepsilon}\big) =- T^{\text{eff}}_{\epsilon\varepsilon}, ~~\text{where}~~T^{\text{eff}}_{\epsilon\varepsilon}=\big(T_{\epsilon\varepsilon}+\beta\,\theta_{\epsilon\varepsilon}\big),~~
\end{eqnarray}
where $f_\mathcal{Q}=\frac{d f}{d \mathcal{Q}}$, and $T_{\epsilon\,\varepsilon}$ is the energy-momentum tensor and extra source $\theta_{\epsilon\,\varepsilon}$, whose forms are
\begin{eqnarray}
\label{eq9}
&& T_{\epsilon\varepsilon}=-\frac{2}{\sqrt{-g}}\frac{\delta\left(\sqrt{-g}\,\mathcal{L}_m\right)}{\delta g^{\epsilon\varepsilon}}\,\\
&& \theta_{\epsilon\,\varepsilon}=-\frac{2}{\sqrt{-g}}\frac{\delta\left(\sqrt{-g}\,\mathcal{L}_\theta\right)}{\delta g^{\epsilon\varepsilon}}, \label{eq10}
\end{eqnarray}
Moreover, From Eq. (\ref{eq1}), we are able to derive the extra constraint over the connection as
\begin{equation}\label{eq11}
\bigtriangledown_\epsilon \bigtriangledown_\varepsilon \left(\sqrt{-g}\,f_Q\,P^\gamma_{\,\,\,\,\epsilon\varepsilon}\right)=0.
\end{equation}
The torsionless and curvatureless constraints render the affine connection as
\begin{equation}\label{eq12}
\Gamma^\lambda_{\,\,\,\epsilon\varepsilon}=\left(\frac{\partial x^\lambda}{\partial\xi^\beta}\right)\partial_\epsilon \partial_\varepsilon \xi^\beta.
\end{equation}
We can make a special coordinate choice, the so-called coincident gauge, so that $\Gamma^\lambda_{\,\,\,\epsilon\varepsilon}=0$. Then, the nonmetricity Eq. (\ref{eq2}) reduces to
\begin{equation}\label{eq13}
\mathcal{Q}_{\lambda\epsilon\varepsilon}=\partial_\lambda g_{\epsilon\varepsilon},
\end{equation}
which vastly simplifies the calculation since only the metric function is the fundamental variable. However, in this case, the action no longer remains diffeomorphism invariant, except for standard General Relativity \cite{Koivisto}. One can use the covariant formulation of $f(\mathcal{Q})$ gravity to avoid such an issue. Since the affine connection in Eq. (\ref{eq12}) is purely inertial, one could use the covariant formulation by first determining affine connection in the absence of gravity \cite{Zhao}. Here, we would like to find gravitationally decoupled solutions for $f(\mathcal{Q})$ gravity describing  compact objects. To this end we consider the standard static spherically symmetric line element of the form,
\begin{equation}\label{eq14}
ds^2=-e^{a(r)}dt^2+e^{b(r)}dr^2+r^2d\theta^2+r^2\text{sin}^2\theta \,d\phi^2,
\end{equation}
Here, where $a(r)$ and $b(r)$ are metric potentials and depend upon the radial distance $r$ which ensures that the spacetime is static. For the current analysis, we are going to work with an anisotropic matter distribution, then the effective energy-momentum tensor $T^{\text{eff}}_{\epsilon\varepsilon}$ can be expressed as:
\begin{eqnarray}
\label{eq15}
&& \hspace{-0.9cm} T^{\text{eff}}_{\epsilon\,\varepsilon}=\left(\rho^{\text{eff}}+p^{\text{eff}}_t\right)u_{\epsilon}\,u_{\varepsilon}-p^{\text{eff}}_t\,g_{\epsilon\,\varepsilon}+\left(p^{\text{eff}}_r-p^{\text{eff}}_t\right)v_{\epsilon}\,v_{\varepsilon},~~
\end{eqnarray}
where $\rho^{\text{eff}}$ and $u_{\epsilon}$ are the effective density and the four-velocity vector, respectively. Besides $v_{\epsilon}$ is the unitary space-like vector in the radial direction, $p^{\text{eff}}_r$ is the effective radial pressure in the direction of $u_{\epsilon}$, and $p^{\text{eff}}_t$ is a effective tangential pressure orthogonal to $v_{\epsilon}$. 
Now, the nonmetricity scalar for the metric (\ref{eq14}) is calculated as:
\begin{equation}\label{eq16}
\mathcal{Q}=-\frac{2 e^{-b(r)} \left(r a'(r)+1\right)}{r^2}, 
\end{equation}

\begin{figure}
    \centering    
     \includegraphics[width=8.7cm,height=6.8cm]{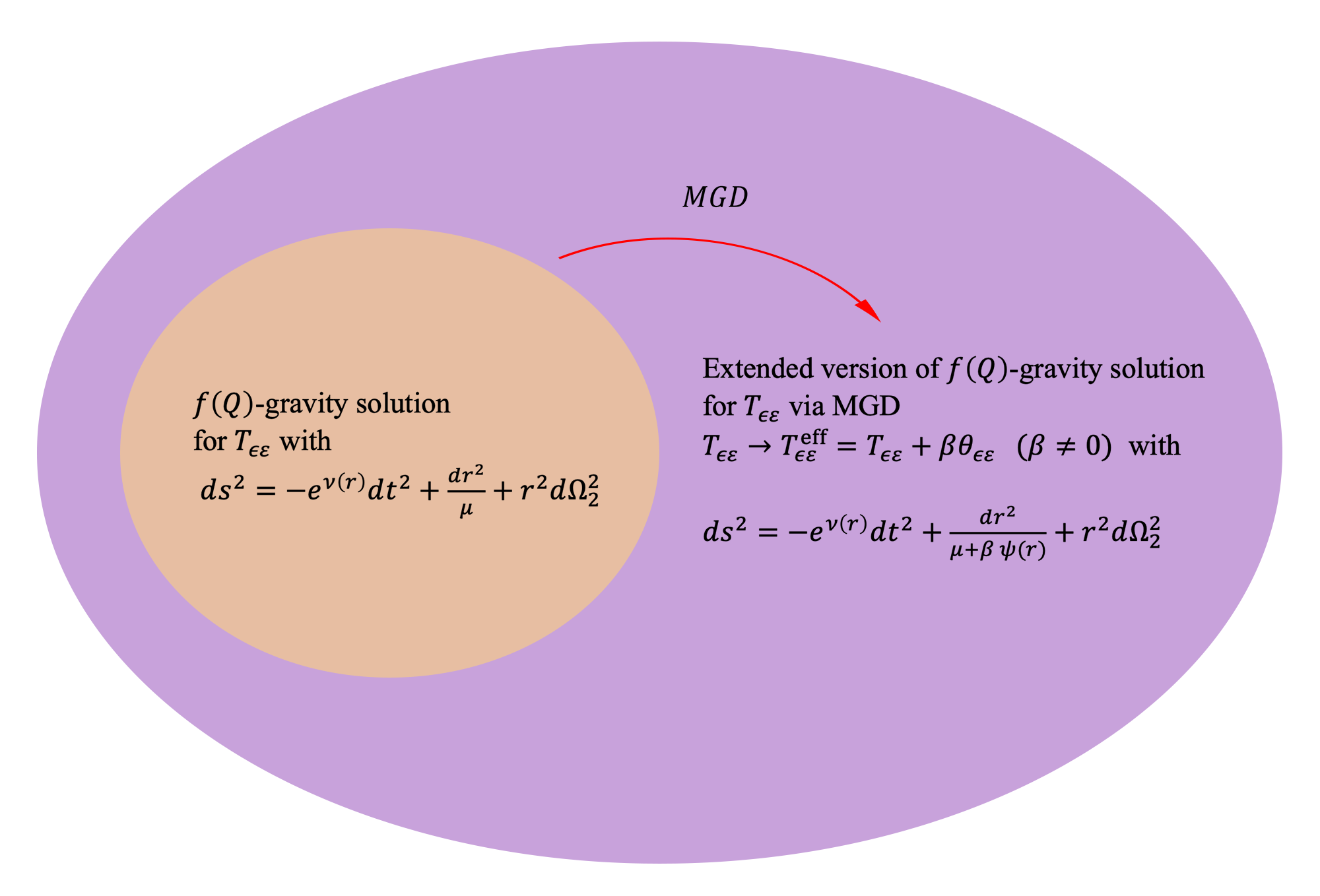}
    \caption{This diagram displays how pure $f(\mathcal{Q})$-gravity solutions can be extended via the MGD formalism to a more generalized form of anisotropic domain.}
    \label{fig1}
\end{figure}

For the anisotropic fluid (\ref{eq17}), the independent components of the equations of motion (\ref{eq8}) in $f(\mathcal{Q})$ gravity are given as,
\begin{eqnarray}
&& \hspace{-0.6cm}\rho^{\text{eff}} =\frac{f(\mathcal{Q})}{2}-f_{\mathcal{Q}}\Big[\mathcal{Q}+\frac{1}{r^2}+\frac{e^{-b}}{r}(a^\prime+b^\prime)\Big],\label{eq17}\\
&& \hspace{-0.6cm} p^{\text{eff}}_r=-\frac{f(\mathcal{Q})}{2}+f_{\mathcal{Q}}\Big[\mathcal{Q}+\frac{1}{r^2}\Big],\label{eq18}\\
&& \hspace{-0.6cm} p^{\text{eff}}_t=-\frac{f(\mathcal{Q})}{2}+f_{\mathcal{Q}}\Big[\frac{\mathcal{Q}}{2}-e^{-b} \Big\{\frac{a^{\prime \prime}}{2}+\Big(\frac{a^\prime}{4}+\frac{1}{2r}\Big)(a^\prime-b^\prime)\Big\}\Big],~~~~\label{eq19}\\
&& \hspace{-0.2cm}0=\frac{\text{cot}\theta}{2}\,\mathcal{Q}^\prime\,f_{\mathcal{Q}\mathcal{Q}}, \label{eq20}
\end{eqnarray}
where $f_{\mathcal{Q}}=\frac{\partial f}{\partial \mathcal{Q}}$. In the background of $f(T)$ theory, it is mentioned that the nonzero off-diagonal metric components of the field equations in $f(T)$ theory, which derives from the specific gauge choice, restricts the functional form of $f(T)$ \cite{rm7}. Therefore, it would also put circumscription on the functional form of $f(\mathcal{Q})$ theory, which effectively provides us a mathematical background for the model building of $f(\mathcal{Q})$ theory. Recently, Wang et al. \cite{Wang} investigated the possible functional forms for $f(\mathcal{Q})$ gravity under the static and spherically symmetric spacetime with an anisotropic fluid. In particular, they have shown that there is no exact Schwarzschild solution for the nontrivial $f(\mathcal{Q})$ function. They also analyzed the deviation of the metric from the exact Schwarzschild solution by considering the nonmetricity scalar $\mathcal{Q}$ being constant. In view of above discussion, we take only $f_{QQ}$ coefficient from the off-diagonal component given in Eq.(\ref{eq20}) to be zero for obtaining the solution of $f(\mathcal{Q})$-gravity which restricts functional form of $f$ as, 
\begin{eqnarray}
&& f_{\mathcal{Q}\mathcal{Q}}=0~~~~~\Longrightarrow~~~~ f(\mathcal{Q})=\alpha_1\,\mathcal{Q}+\alpha_2 ,  \label{eq21}
\end{eqnarray}
where $\alpha_1$ and $\alpha_2$ are constants. By plugging of Eqs.(\ref{eq16}) and (\ref{eq21}), the Eq.(\ref{eq17})-(\ref{eq19}) provides the following explicit form of equations of motion,
\begin{eqnarray}
&& \hspace{-0.4cm}\rho^{\text{eff}} = \frac{1}{2r^2} \Big[2 \alpha_1 -r^2 \alpha_2 +2 e^{-b} \alpha_1  \left(r\, b^\prime-1\right)\Big],\label{eq22}\\
&& \hspace{-0.4cm} p^{\text{eff}}_r=\frac{1}{2r^2} \Big[-2 \alpha_1 +r^2 \alpha_2 +2 e^{-b} \alpha_1  \left(r\, a^\prime+1\right)\Big],\label{eq23}\\
&& \hspace{-0.4cm} p^{\text{eff}}_t=\frac{e^{-b}}{4r} \Big[  2 e^{b} r \alpha_2 +\alpha_1  \left(2+r a^\prime\,\right) \left(a^\prime-b^\prime\right)+2 r \alpha_1  a^{\prime \prime}\Big],~~~~~~~\label{eq24} 
\end{eqnarray}
We note that the covariant derivative of effective energy-momentum tensor under the assumption of spherical symmetry (\ref{eq14}) vanishes i.e. $\bigtriangledown^\epsilon T^{\text{eff}}_{\epsilon\,\varepsilon}=0$, which gives  
\begin{eqnarray}
-\frac{a^\prime}{2}(\rho^{\text{eff}}+p^{\text{eff}}_r)-(p^{\text{eff}}_r)^{\prime}+\frac{2}{r}( p^{\text{eff}}_{t}-p^{\text{eff}}_r)=0.\label{eq25}
\end{eqnarray}
The above Eq.(\ref{eq25}) is known as a Tolman-Oppenheimer-Volkoff (TOV) equation in $f(\mathcal{Q})$-gravity under the linear functional form of $f(\mathcal{Q})$ for the Eq.(\ref{eq8}). The TOV equation in $f(\mathcal{Q})$-gravity is similar to the conservation equation in classical general relativity.  Our next strategy is to find an exact solution of the field equations (\ref{eq22})-(\ref{eq24}) describing a strange star model. We note that the system of field equations are highly non-linear and hence we are faced with a difficult task of solving them.  Therefore, we apply another approach known as gravitational decoupling through minimal geometric deformation (MGD) technique under a particular transformation along the gravitational potential, 
\begin{eqnarray}
&& a(r) \longrightarrow \nu(r)+\beta\, \xi(r), \label{eq26}\\
&& e^{-b(r)} \longrightarrow \mu(r)+\beta\, \psi(r)  \label{eq27}
\end{eqnarray}
where $\xi(r)$ and  $\psi(r)$ are called the geometric deformation functions or decoupling functions along the temporal and radial metric components, respectively. This deformation can be set suitably through the decoupling constant $\beta$. As usual, when $\beta =0$, the standard $f(\mathcal{Q})$-gravity theory is recovered. Since we are applying here the MGD approach which allow us to set  $ \xi(r) = 0$ and  $\psi(r)\ne 0$. This indicates that the suitable transformation acts along only the radial component of the metric function and the temporal component is unaffected. A schematic
diagram has been shown in \ref{fig1}, where the $f(\mathcal{Q})$-gravity solution is extended to be a solution in the new gravitational sector using the MGD approach. This MGD technique divides the decoupled system (\ref{eq22})-(\ref{eq24}) into two subsystems. The first system corresponds to  $T_{\epsilon\,\varepsilon}$ and other system for the extra source $\theta_{\epsilon\,\varepsilon}$. In order to write the first system, we consider the energy-momentum tensor ${T}_{\epsilon\,\varepsilon}$ that describes an anisotropic matter distribution given by, 
\begin{equation}\label{eq28}
T_{\epsilon\,\varepsilon}=\left(\rho+p_t\right)u_{\epsilon}\,u_{\varepsilon}+p_t\,\delta_{\epsilon\,\varepsilon}+\left(p_r-p_t\right)v_{\epsilon}\,v_{\varepsilon},
\end{equation}
where, $\rho$ denotes a energy density while $p_{r}$ and $p_{t}$ denote the radial pressure and tangential pressure, respectively for the seed solution. Then, the effective quantities can be written as,
\begin{eqnarray}
\rho^{\text{eff}}=\rho+\beta \,\theta^0_0,~~p^{\text{eff}}_r=p_r-\beta\,\theta^1_1,~~p^{\text{eff}}_t=p_t-\beta\,\theta^2_2.~~~~  \label{eq29}
\end{eqnarray}
and corresponding effective anisotropy is,
\begin{eqnarray} 
&&\hspace{-0.7cm} \Pi^{\text{eff}}=p^{\text{eff}}_t-p^{\text{eff}}_r= \Pi+\Pi_{\theta},
 \label{eq30}\\
&&\hspace{-0.7cm} \text{where},~~~~\Pi= p_t-p_r~~~~~\text{and}~~~~~\Pi_\theta= \beta (\theta^1_1-\theta^2_2)\nonumber. 
\end{eqnarray}
We observe that effective anisotropy is the sum of two anisotropies corresponding to $T_{\epsilon\,\varepsilon}$ and $\theta_{\epsilon\,\varepsilon}$.  The anisotropy $\Pi_{\theta}$ is generated by gravitational decoupling that may enhance the effective anisotropy but this will solely depend on the behavior of $\Pi_{\theta}$. Now applying the transformations (\ref{eq26}) and (\ref{eq27}), the system  (\ref{eq22})-(\ref{eq24}) yields the following set of equations depending on the gravitational potentials $\nu$ and $\mu$, (i.e. when $\beta = 0$) as,
\begin{eqnarray}
&&\hspace{-0.8cm}\rho= \frac{\alpha_1 }{r^2}-\frac{\mu \alpha_1 }{r^2}-\frac{\mu^{\prime} \alpha_1 }{r}-\frac{\alpha_2 }{2},\label{eq31}\\
&&\hspace{-0.8cm} p_r=-\frac{\alpha_1 }{r^2}+\frac{\mu \alpha_1 }{r^2}+\frac{\nu^{\prime} \mu \alpha_1 }{r}+\frac{\alpha_2 }{2}, \label{eq32}\\
&&\hspace{-0.8cm} p_t=\frac{\mu^{\prime} \nu^{\prime} \alpha_1 }{4}+\frac{\nu^{\prime \prime} \mu \alpha_1 }{2}+\frac{\nu^{\prime 2} \mu \alpha_1}{4}  +\frac{\mu^{\prime} \alpha_1 }{2 r}+\frac{\nu^{\prime} \mu \alpha_1 }{2 r}+\frac{\alpha_2 }{2},~~~~~~\label{eq33}
\end{eqnarray}
where the Eq.(\ref{eq25}) leads to the following,
\begin{eqnarray}
-\frac{\nu^\prime}{2}(\rho+p_r)-(p_r)^{\prime}+\frac{2}{r}( p_{t}-p_r)=0,~~\label{eq34}
\end{eqnarray}
which is a TOV equation for the system (\ref{eq31})-(\ref{eq33}) whose solution can be given by the following spacetime, 
\begin{equation}\label{eq35}
ds^2=-e^{\nu(r)}dt^2+\frac{dr^2}{\mu(r)}+r^2d\theta^2+r^2\text{sin}^2\theta \,d\phi^2,
\end{equation}
The other system of equations can be obtained by turning on $\beta$ as,
\begin{eqnarray}
&&\hspace{-0.7cm}\theta^{0}_0=-\alpha_1 \Big(\frac{\psi   }{r^2}+\frac{\psi^\prime }{r}\Big), \label{eq36}\\
&&\hspace{-0.7cm}\theta^1_1=-\alpha_1\Big(\frac{\psi  }{r^2}+\frac{\nu^{\prime} \psi   }{r}\Big), \label{eq37}\\
&&\hspace{-0.7cm}\theta^2_2=-\alpha_1 \Big(\frac{1}{4} \psi^\prime \nu^{\prime}   +\frac{1}{2} \nu^{\prime \prime} \psi   +\frac{1}{4} \nu^{\prime 2} \psi  +\frac{\psi^\prime  }{2 r}+\frac{\nu^{\prime} \psi  }{2 r}\Big), \label{eq38}
\end{eqnarray}
The linear combination of the equations (\ref{eq36})-(\ref{eq38}) provide the following relation,
\begin{eqnarray}
-\frac{\nu^{\prime}}{2} (\theta^0_0-\theta^1_1)+ (\theta^1_1)^\prime+\frac{2}{r} (\theta^1_1-\theta^2_2)=0, \label{eq39}
\end{eqnarray}
The mass function corresponding to both systems can be written as,
\begin{eqnarray}
&&\hspace{-0.7cm} m_{Q}=\frac{1}{2} \int^r_0 \rho(x)\, x^2 dx~\text{and}~m_{\phi}= \frac{1}{2}\,\int_0^r \theta^0_0 (x)\, x^2 dx, \label{eq40}
\end{eqnarray}
where $m_{Q}(r)$ and $m_{\phi}(r)$ denote the mass functions corresponding to the sources $T_{\epsilon \varepsilon}$ and $\theta_{\epsilon \varepsilon}$, respectively.

Now the advantage of MGD-decoupling becomes somewhat clear such as:  we can extend any known solutions associated with the action $\mathcal{S}_{Q} $ with solution $\{T_{ \epsilon\,\varepsilon}, \nu, \mu\}$ of the system given by Eqs.(\ref{eq31})-(\ref{eq33}) into the domain of beyond modified gravity denoted by action $\mathcal{S}$ whose equation of motions are displayed in  Eqs.(\ref{eq22})-(\ref{eq24}), by solving the unconventional gravitational system of equations Eqs.(\ref{eq36})-(\ref{eq38}) to determine $\{\theta_{ \epsilon\,\varepsilon }$, $\xi$, $\psi\}$. Hence we can generate the "$\theta$-version" of any $\{T_{ \epsilon\,\varepsilon}, \nu, \mu\}$-solution as 
\begin{eqnarray}
\{T_{ \epsilon\,\varepsilon},~ \nu(r),~ \mu(r)\} \Longrightarrow \{T^{\text{eff}}_{ \epsilon\,\varepsilon}, ~a(r),~ b(r)\}
\end{eqnarray}
The above relation describes a straightforward way to study the consequences of beyond the symmetric teleparallel gravity.

\section{Minimally deformed solution for strange star (SS) in $f(\mathcal{Q})$ gravity} \label{sec3}
In this section, we will discuss the solution of both systems (\ref{eq31})-(\ref{eq34}) and (\ref{eq36})-(\ref{eq39}) for strange star model. For this purpose, we assume that the internal structure of the model for the seed spacetime corresponding to $T_{\epsilon \varepsilon}$ is composed of strange quarks which can be described by the  MIT bag equation of state (EOS) \cite{Chodos:1974}. The MIT bag model describes degenerated Fermi gas of quarks up ($u$), down ($d$) and strange ($s$) \cite{Chodos:1974,Farhi:1984} which is given by a non-interacting EoS corresponding to the matter variables $\rho$ and $p_r$ as,
\begin{eqnarray} \label{eq41}
p_r=\sum_{f} p^{f}-\mathcal{B}_g,~~~~f=u,~d,~s
\end{eqnarray}
here ${p^f}$ is called the individual pressures corresponding to $(u)$, $(d)$ and $(s)$ quark flavor which is neutralized by the total external Bag pressure or Bag constant
$\mathcal{B}_g$, and then the energy density  ($\rho$) for the deconfined quarks interior related to the MIT Bag model can be written as, 
\begin{eqnarray}
\rho=\sum_{f}\rho^{f}+\mathcal{B}_g,~~~\text{where}~~~\rho^f=3p^f.\label{eq42}
\end{eqnarray}
Using the equations (\ref{eq41}) and (\ref{eq42}) with the relation $\rho^f=3p^f$, we can write the explicit form of the MIT bag equation of state (EOS) for strange quark stars as,
\begin{eqnarray} \label{eq43}
p_r=\frac{1}{3}(\rho-4\mathcal{B}_g).
\end{eqnarray} 
Now using the Eqs.(\ref{eq31}) and (\ref{eq32}) with EoS (\ref{eq43}), we find the following differential equation in terms of the metric functions  $\mu(r)$ and $\nu(r)$ as, 
\begin{eqnarray}
&&\hspace{-0.7cm} \alpha_1\,(4 \mu+\mu^{\prime} r+3 \nu^{\prime} \mu\, r-4)  +2 r^2 \alpha_2+ 4\,r^2\, \mathcal{B}_g=0, \label{eq44}
\end{eqnarray}
To solve the above equation, we propose the potential ansatz corresponding to the Tolman IV ansatz for the metric function $\mu(r)$ of the form,

\begin{eqnarray}
\mu(r)={1 \over (1+Ar^2+Br^4)},  \label{eq45}
\end{eqnarray}

where $A$ and $B$ are constants with dimensions $\text{length}^{-2}$ and $\text{length}^{-4}$, respectively. By substituting $\mu$ from Eg.(\ref{eq45}) into Eq.(\ref{eq44}) and integrating, we obtain 
\begin{eqnarray}
&&\hspace{-0.5cm} \nu(r)=-\frac{1}{18\,\alpha_1}\Big[2 B r^6 (2 \mathcal{B}_g + \alpha_2) + 6 r^2 (2 \mathcal{B}_g - 2 A \alpha_1 + \alpha_2) \nonumber\\&&\hspace{0.5cm}+ 
 3 r^4 ( A (2 \mathcal{B}_g + \alpha_2)-2 B \alpha_1) - 
 6 \alpha_1  \ln[1 + A r^2 + B r^4]\nonumber\\&&\hspace{0.5cm} + 18\alpha_1\,C, \label{eq46}
\end{eqnarray}
where $C$ is a constant of integration. Since now we have the potentials $\mu(r)$ and $\nu(r)$, then we find the expressions for $\rho$, $p_r$, and $p_t$ as,
\begin{eqnarray}
&&\hspace{-0.5cm} \rho(r)= \frac{\left[3 A+\left(A^2+5 B\right) r^2+2 A B r^4+B^2 r^6\right] \alpha_1 }{\left(1+A r^2+B r^4\right)^2}-\frac{\alpha_2 }{2},\\
&&\hspace{-0.5cm}  p_r(r) =\frac{1}{6 (1 + A r^2 + B r^4)^2} \big[2 \big\{3 A+\left(A^2+5 B\right) r^2+2 A B r^4\nonumber\\&&\hspace{0.6cm}+B^2 r^6\big\} \alpha_1 -\big(1+A r^2+B r^4\big)^2 \alpha_2-8 \mathcal{B}_g \big(1+A r^2\nonumber\\&&\hspace{0.6cm}+B r^4\big)^2 \big],\\
&&\hspace{-0.5cm} p_t(r)=\frac{1}{18(1+A r^2+B r^4)^3 \alpha_1}\Big[8 \mathcal{B}^2_g r^2 \left(1+A r^2+B r^4\right)^4\nonumber\\&&\hspace{0.5cm}+60 B r^2 \alpha_1 ^2+44 B^2 r^6 \alpha_1 ^2+56 B^3 r^{10} \alpha_1 ^2+8 B^4 r^{14} \alpha_1 ^2\nonumber\\&&\hspace{0.5cm}-3 \alpha_1  \alpha_2  -37 B r^4  \alpha_1  \alpha_2 -73 B^2 r^8 \alpha_1  \alpha_2 -47 B^3 r^{12} \nonumber\\&&\hspace{0.5cm} \times\alpha_1  \alpha_2 -8 B^4 r^{16} \alpha_1  \alpha_2 +2 r^2 \alpha_2 ^2+8 B r^6 \alpha_2 ^2+12 B^2 r^{10} \nonumber\\&&\hspace{0.5cm} \times \alpha_2 ^2+8 B^3 r^{14} \alpha_2 ^2 +2 B^4 r^{18} \alpha_2 ^2+2 A^4 r^6  \left(r^2 \alpha_2-2 \alpha_1 \right)^2\nonumber\\&&\hspace{0.5cm}-4 \mathcal{B}_g (1 +A r^2+B r^4)^2 \big\{\big(6+4 A^2 r^4+20 B r^4+4 B^2 r^8\nonumber\\&&\hspace{0.5cm} +A r^2 (15 +8 B r^4)\big) \alpha_1  -2r^2 (1+A r^2+B r^4)^2 \alpha_2 \big\}\nonumber\\&&\hspace{0.5cm}+A^3 r^4 \big(4 \left(9+8 B r^4\right) \alpha_1 ^2-r^2 \left(37+32 B r^4\right) \alpha_1  \alpha_2 \nonumber\\&&\hspace{0.5cm} +8 r^4 (1+B r^4) \alpha_2 ^2\big) +p_{t1}(r)\Big],
\end{eqnarray}

Now we have completely solved the first system. Then we focus on second set of equations (\ref{eq36})-(\ref{eq38}) corresponding to $\theta$-sector which depends on the deformation function $\psi(r)$.
First, we would like to highlight some essential features for $\theta$-system given by equations
(\ref{eq36})-(\ref{eq38}).  If we look at this system of equations, we found that it is very similar to the standard
spherically symmetric field equations for anisotropic matter distribution with energy-momentum tensor $\theta_{\epsilon\,\varepsilon}$; $\{ \rho(r)=\theta^0_0(r),~~ p_r(r)=-\theta^1_1(r),~~ p_t(r)=-\theta^2_2(r)\}$ as well as and its corresponding conservation equation.  Nevertheless, it cannot be exactly obtained as the $f(\mathcal{Q})$-gravity field equations for spherically symmetric spacetime with radial metric component $\psi(r)$ because the geometrical part of Eqs. (\ref{eq36}) -(\ref{eq38}) are not similar the standard expressions for energy momentum tensor $T_{\epsilon\,\varepsilon}$  due to the missing term $\frac{\alpha_1}{r^2}$ and $\frac{\alpha_1}{r^2}$.
 Regardless of the above, the system (\ref{eq36})-(\ref{eq39}) may be formally obtained similar as the standard gravity field equations in $f(\mathcal{Q})$-gravity for an anisotropic system with energy-momentum tensor $\theta^{\ast}$ defined as,
 \begin{eqnarray}
 \theta^{\ast \epsilon}_{\varepsilon}= \theta^{\epsilon}_{\varepsilon} +\Big(\frac{\alpha_1}{r^2}-\frac{\alpha_2}{2}\Big) \big(\delta^0_\epsilon\,\delta^\varepsilon_0+\delta^1_\epsilon\,\delta^\varepsilon_1+\delta^2_\epsilon\,\delta^\varepsilon_2)
 \end{eqnarray}
 and corresponding conservation equation 
 \begin{eqnarray}
-\frac{\nu^{\prime}}{2} (\theta^{\ast 0}_0-\theta^{\ast 1}_1)+ (\theta^{\ast 1}_1)^\prime+\frac{2}{r} (\theta^{\ast 1}_1-\theta^{\ast 2}_2)=0, \label{eq391}
\end{eqnarray}
and the corresponding spacetime
\begin{equation}\label{eq53}
ds^2=-e^{\nu(r)}dt^2+\frac{dr^2}{\psi(r)}+r^2d\theta^2+r^2\text{sin}^2\theta \,d\phi^2,
\end{equation}
It is evident that we have three independent equations with four unknowns. Therefore, we need only one additional information to close the $\theta$-system. We have  only two alternative approaches to solve the $\theta$-sector as either taking a particular viable functional form of $\psi(r)$, or linear equation of state (EoS) between the $\theta_{\epsilon \varepsilon}$. But, it is not a trivial task to find physically acceptable solution for the system. On the other hand, we should keep in our mind that the deformation function $\psi(r)$ must vanish at centre (since for any physically acceptable model $e^{-b(0)}=\mu(0)=1$, which implies $f(0)=0$). Apart from both approaches, we
can take advantage of the shape of the system of equations (\ref{eq36}) and (\ref{eq37}) for the source ($\theta_{\epsilon \varepsilon}$) $\theta^{\ast \epsilon}_{\varepsilon}$. Hence, we can take the following choices of $\psi(r)$ in the terms of known metric functions $\mu(r)$ and $\nu(r)$ (when $\alpha_2=0$),
\begin{eqnarray}
\psi(r)=\mu(r)-1~~\mbox{and}~~\psi(r)=\frac{1}{ 1+r\,\nu'(r)}- \mu(r).~~~~
\end{eqnarray}
As we can see from above equation that $\psi(r)$ is free from singularity and $\psi(0)=0$. As a consequence, these forms of $\psi(r)$ allow us to mimic the $\theta^0_0$ with 
energy density $\rho$ and mimic the $\theta^1_1$ with the radial pressure $p_r$. This approach is very popular amongst researchers which has been widely used to solve $\theta$-sector for compact stellar models in different modified gravity theories \cite{sunil5d1,sunil5d3,sunil5d4,sunil5d2,sharif1,sharif2,sharif3}. Due to its success in modeling compact objects in both classical GR and modified gravity theories, we employ the mimic approach in the next section to solve the system of equations (\ref{eq36})-(\ref{eq38}).  

\subsection{\textbf{Mimicking of the density constraint i.e. $\rho(r)=\theta^0_0(r)$}} \label{sec3.1}
By mimicking of seed density $\rho$ from Eq.(\ref{eq31}) with the  $\theta$-component $\theta^0_0$ via Eq.(\ref{eq36}), we obtain a differential equation in $\psi(r)$ as,
\begin{eqnarray}
2(1+\psi-\mu+r\,\psi^{\prime}-r\,\mu^{\prime}) \alpha_1 -r^2\,\alpha_2 =0,
\end{eqnarray}
Now by inserting the gravitational potentials $\mu(r)$ and $\nu(r)$ from Eq.(\ref{eq45}) into Eq.(\ref{eq46}) and integrating yields 
\begin{eqnarray}
\psi(r)=\frac{(1+A r^2+B r^4) (r^2 \alpha_2-6  \alpha_1)+6 \alpha_1 }{6 \alpha_1\,(1+A r^2+B r^4)},\label{eq50}
\end{eqnarray}
where we have put the arbitrary constant of integration to be zero in order to ensure the non-singular nature of $\psi(r)$ at the stellar centre. Using the expression (\ref{eq50}), the components for the $\theta$-sector are obtained as,
\begin{eqnarray}
&&\hspace{-0.3cm}\theta^0_0(r)=\frac{10 B r^2 \alpha_1 + 2 B^2 r^6 \alpha_1 - \alpha_2 - 2 B r^4 \alpha_2+\theta_{11}(r)}{2 (1+A r^2+B r^4)^2},~~~~~~~~\\
&&\hspace{-0.3cm}\theta^1_1(r)=\frac{\alpha_2 -(6 A +6 B r^2) \alpha_1 +(A r^2 +B r^4)\alpha_2 }{18 \alpha_1 (1 + A r^2 + B r^4)^2 [\theta_{12}(r)]^{-1}},~~~
\end{eqnarray}
We avoid to write here the $\theta^2_2(r)$ due to its cumbersome expression. 
\subsection{\textbf{Mimicking of the pressure constraint i.e. $p_r(r)=\theta^1_1(r)$}} \label{sec3.2}
In this approach, we mimic the radial pressure $p_r$ corresponding to seed solution with its $\theta$-component $\theta^1_1$ by using the known $\mu(r)$ and $\nu(r)$,  we find directly the expression for decoupling function $\psi(r)$ as,
\begin{eqnarray}
\psi(r)=\frac{r^2 \big[8 \mathcal{B}_g (1+A r^2+B r^4)^2+\psi_1(r) \big]}{2 \left(1+A r^2+B r^4\right) \,\psi_2(r)},
\end{eqnarray}
where,
\begin{eqnarray}
&&\hspace{-0.3cm} \psi_1(r)=\left(1+A r^2+B r^4\right)^2 \alpha_2-2[3 A+\left(A^2+5 B\right) r^2\nonumber\\ &&\hspace{0.7cm}+2 A B r^4+B^2 r^6] \alpha_1\,\nonumber\\
&&\hspace{-0.3cm} \psi_2(r)=4 \mathcal{B}_g (r+A r^3+B r^5)^2-(3+9 A r^2+4 A^2 r^4\nonumber\\ &&\hspace{0.7cm}+11 B r^4+8 A B r^6+4 B^2 r^8) \alpha_1 -2r^2 (1+A r^2\nonumber\\ &&\hspace{0.7cm}+B r^4)^2 \alpha_2. \nonumber
\end{eqnarray}
Then the $\theta$-components for this solution are, 
\begin{eqnarray}
&&\hspace{-0.5cm} \theta^0_0(r)=\frac{\alpha_1 \, \theta_{21}(r)}{2 (1+A r^2+B r^4)^2\, \theta_{22}(r)},\\
&&\hspace{-0.5cm} \theta^1_1(r)= \frac{8 \mathcal{B}_g \,(1+A r^2+B r^4)^2-10 B r^2 \alpha_1 +\theta_{23}(r)}{6 \left(1+A r^2+B r^4\right)^2},~~~~~~~~
\end{eqnarray}
The $\theta^2_2(r)$ is long and complicated so we chose not to present it here.  
\section{Exterior space--time and Junctions conditions} \label{sec4}
In this section, we will discuss the suitable boundary conditions  for the obtained minimally deformed solutions by matching of the interior and exterior spacetimes at the surface of the object. Following recent work \cite{Wang}, the suitable exterior spacetime in $f(\mathcal{Q})$ gravity is described by the Schwarzschild Anti-de Sitter spacetime,
\begin{eqnarray}
&&\hspace{-0.0cm} ds^2_+ =-\bigg(1-\frac{2{\mathcal{M}}}{r}-{\Lambda r^2 \over 3}\bigg)\,dt^2+ \bigg(1-\frac{2{\mathcal{M}}}{r}-{\Lambda r^2 \over 3}\bigg)^{-1} \nonumber\\
&&\hspace{0.5cm} \times dr^2+r^2 \Big(d\theta^2  +\sin^2\theta\,d\phi^2 \Big),~\label{eq57}
\end{eqnarray} 
where, $\mathcal{M}$ is the total mass and $\Lambda$ the cosmological constant. The interior region ($0 \le r \le R$) of the self-gravitating system can be given by the deformed spacetime,
\begin{eqnarray}
&&\hspace{-0.5cm} ds^2_{-}= -e^{\nu(r)}\,dt^2+ \big[\mu(r)+\beta\,\psi(r)\big]^{-1} dr^2+r^2(d\theta^2\nonumber\\
&&\hspace{3.5cm}+\sin^2\theta\,d\phi^2),~~~ \label{eq58}
\end{eqnarray} 
where the interior mass function for minimally deformed spacetime in $f(\mathcal{Q})$ gravity can be cast as,
\begin{eqnarray}
\hat{m}_Q(r)=m_Q(r)-\frac{\alpha_1\,\beta}{2}\,r\,\psi(r), \label{eq59}
\end{eqnarray}
where the mass function ($m_Q(r)$) is given by Eq.(\ref{eq40}). Furthermore, the line elements (\ref{eq57}) and (\ref{eq58}) must satisfy the Israel-Darmois matching conditions \cite{Israel,Darmois} at the boundary of the star ($r=R$), which ensures the continuity of the metric potentials across the boundary $r=R$ and the vanishing of the radial pressure. The matching of metric potentials across the boundary $r=R$ implies,
\begin{eqnarray}
e^{a^{-}(r)}|_{r=R}=e^{a^{+}(r)}|_{r=R}~\mbox{and}~e^{b^{-}(r)}|_{r=R}=e^{b^{+}(r)}(r)|_{r=R},\nonumber \\
\label{eq60}
\end{eqnarray}
Using the spacetimes (\ref{eq57}) and (\ref{eq58}), the conditions (\ref{eq60}) gives,
\begin{eqnarray}
&&\hspace{-0.5cm} e^{a(R)}=e^{\nu(R)}= 1-\frac{2{\mathcal{M}}}{R}-{\Lambda \over 3}~R^2, \label{eq61}\\
&&\hspace{-0.5cm} e^{-b(R)}=\mu(R)+\beta\,\psi(R)=1-\frac{2{\mathcal{M}}}{R}-{\Lambda \over 3}~R^2, \label{eq62}
\end{eqnarray}
Then $\mathcal{M}=\hat{M}_Q/\alpha_1$, and $\Lambda=\alpha_2/2\alpha_1$, where $\hat{m}_Q(R)=\hat{M}_Q$. It is clearly observed that when $\alpha_1=1$ and $\alpha_2=0$, the Schwarzschild Anti-de Sitter spacetime (\ref{eq57}) reduces into Schwarzschild exterior solution. Similarly, the extrinsic curvature (or second fundamental form) of spheres,
\begin{eqnarray}
K_{\epsilon\,\varepsilon}=\bigtriangledown_{\epsilon}\,r_{\varepsilon}, \label{eq63}
\end{eqnarray}
where, $r_{\varepsilon}$ denotes the unit radial vector normal to the surface $\Sigma$. By taking above Eq.(\ref{eq63}) together with equation (\ref{eq15}), we obtain the second fundamental form in terms of effective energy-momentum tensor as,
\begin{eqnarray}
\big[T^{\text{eff}}_{\epsilon\,\varepsilon}\,r^{\varepsilon}\big]_{\Sigma}=\big[(T_{\epsilon\,\varepsilon}+\beta\,\theta_{\epsilon\,\varepsilon})\,r^{\varepsilon}\big]_{\Sigma}=0, \label{eq65}
\end{eqnarray}
which gives,
\begin{eqnarray}
\big[p_r(r)-\beta \theta^1_1 (r)\big]_{\Sigma}=0. \label{eq66}
\end{eqnarray}
The final form of the above matching condition is given as,
\begin{eqnarray}
p_{r}(R)-\beta (\theta^1_1)^{-} (R)=-\beta (\theta^1_1)^{+} (R), \label{eq67}
\end{eqnarray}
The condition (\ref{eq67}) denotes a general second fundamental form related to the Einstein's equation (\ref{eq8}) and energy momentum tensor (\ref{eq15}). While, $(\theta^1_1)^{-}(R)$ and $(\theta^1_1)^{+}(R)$ denote the components of the extra source $\theta_{\epsilon \varepsilon}$ for the internal and external spacetime at $r=R$, respectively. Using the interior geometry of decoupling source $\theta^1_1$ from Eq.(\ref{eq37}), the Eq.(\ref{eq67}) leads to,
\begin{eqnarray}
p_r(R)+\beta\,\alpha_1\,\psi_{_\Sigma} \,\Big[\frac{1 }{R^2}+\frac{\nu^{\prime}_{_\Sigma}   }{R}\Big]=-\beta (\theta^1_1)^{+} (R),\label{eq68} ~~
\end{eqnarray}
where, $\psi_{_\Sigma}=\psi(R)$ and $\nu^{\prime}_{_\Sigma}=\nu^{\prime} (R)$. Furthermore, plugging of the Eq.(\ref{eq37}) for the exterior geometry in the above equation (\ref{eq68}), we get
\begin{eqnarray}
p_r(R)+\beta\,\alpha_1\,\psi_{_\Sigma} \,\Big[\frac{1 }{R^2}+\frac{\nu^{\prime}_{_\Sigma}   }{R}\Big]=\beta\,\alpha_1\,\psi^{\ast}_{_\Sigma} \,\Big[\frac{1 }{R^2}\nonumber\\+\frac{\frac{2\mathcal{M}}{R^2}-\frac{2\Lambda}{3} R   }{R\Big(1-\frac{2{\mathcal{M}}}{R}-{\Lambda \over 3}~R^2\Big)}\Big], ~~
\end{eqnarray}
Here, $\psi^{\ast}_{_\Sigma}=\psi^{\ast}(R)$ denotes a decoupling function corresponding to the exterior space--time at $r=R$ induced by the source $\theta_{\epsilon \varepsilon}$, which can be determined from the spacetime,
\begin{eqnarray}
&&\hspace{-0.5cm} ds^2_+ =-\bigg(1-\frac{2{\mathcal{M}}}{r}-{\Lambda \over 3}~r^2\bigg)\,dt^2+ \bigg[1-\frac{2{\mathcal{M}}}{r}-{\Lambda \over 3}~r^2\nonumber\\
&&\hspace{0.5cm}+\beta\,\psi^{\ast}(r)\bigg]^{-1} dr^2+r^2 \Big(d\theta^2  +\sin^2\theta\,d\phi^2 \Big),~\label{eq69}
\end{eqnarray} 
The Eqs.(\ref{eq61}), (\ref{eq62}), and (\ref{eq69}) are the necessary and sufficient conditions for matching of the interior MGD metric (\ref{eq58}) and exterior ``vacuum" static and spherically symmetric space--time (\ref{eq57}) at the boundary of the star.  Since here we considered the contributions for the new source $\theta_{\epsilon \varepsilon}$ are confined within the stellar interior only and the exterior geometry is given by the exact Schwarzschild Anti-de Sitter solution. Then we must put $\psi^{\ast}_{_\Sigma}=0$ in Eq. (\ref{eq69}). Hence, we find the final form of the condition by substituting $\Phi^{\ast}_{\Sigma}=0$ in Eq. (\ref{eq69}) as, 
\begin{eqnarray}
p_r(R)+\beta\,\alpha_1\,\psi_{_\Sigma} \,\Big[\frac{1 }{R^2}+\frac{\nu^{\prime}_{_\Sigma}   }{R}\Big]=0. \label{eq70}
\end{eqnarray}
The condition (\ref{eq70}) shows some important information regarding the self-gravitating
system:the condition (\ref{eq70}) must be satisfied in order to be consistently coupled with the Schwarzschild Anti-de Sitter geometry (\ref{eq66}) which shows that the effective radial pressure $p^{\text{eff}}_r$ must vanish at the boundary of star $r=R$. However, if the decoupling function  $\psi(r < R)$ is positive with $\beta>0$ yields a weaker gravitational
field due to low mass, [see Eq. (\ref{eq59})]. 
Now using the conditions (\ref{eq61}), (\ref{eq62}) and (\ref{eq70}), we find the constants $A$, $B$ and $\mathcal{M}$ for both solutions (\ref{sec3.1}) and (\ref{sec3.2}) as:
\subsection{\textbf{Constants for solution \ref{sec3.1}}}
\begin{eqnarray}
&&\hspace{-0.5cm}\mathcal{B}_g= \frac{\alpha_1 \left[\mathcal{B}_{g1}(R)+3 \alpha_2 \right]}{4 \left[\alpha_2  \beta R^2 \left(A R^2+B R^4+1\right)-6 \alpha_1 \mathcal{B}_{g2}(R)\right]},\\
&&\hspace{-0.5cm} \mathcal{M}=\frac{R^3 \left[2 \alpha_1\mathcal{M}_1(R)-\alpha_2  \beta \left(A R^2+B R^4+1\right)\right]}{12 \alpha_1 \left(A R^2+B R^4+1\right)},\\
&&\hspace{-0.5cm} C =\frac{1}{18 \alpha_1} \big[C_{11}(R)-6 \alpha_1 \log \left(A R^2+B R^4+1\right)\big]
\end{eqnarray}
\subsection{\textbf{Constants for solution \ref{sec3.2}}}
\begin{eqnarray}
&&\hspace{-0.0cm}\mathcal{B}_g=\frac{1}{8 \left(A R^2+B R^4+1\right)^2}\big[A^2 \left(2 \alpha_1  R^2-\alpha_2  R^4\right)+\alpha_1  A \nonumber\\&&\hspace{0.5cm}\times \left(4 B R^4+6\right)-2 A \alpha_2  R^2 \left(B R^4+1\right)-\alpha_2 +B^2 \nonumber\\&&\hspace{0.5cm}\times \left(2 \alpha_1  R^6-\alpha_2  R^8\right)-2 B \left(\alpha_2  R^4-5 \alpha_1 R^2\right)\big],\\
&&\hspace{-0.0cm} \mathcal{M}=\frac{1}{2} R \left[\frac{A R^2+B R^4}{A R^2+B R^4+1}-\frac{\text{G1} R^2}{3}+\mathcal{M}_2(R)\right],\\
&&\hspace{-0.0cm} C= \ln \left[\frac{2-C_{21}(R)}{2 \left(A R^2+B R^4+1\right)}\right]+\frac{1}{18 \alpha_1 }\big[3 R^4 \{A (\alpha_2 +2 \mathcal{B}_g)\nonumber\\&&\hspace{0.5cm} -2 \alpha_1  B\}-6 \alpha_1  \ln \left(A R^2+B R^4+1\right)+6 R^2 (-2 \alpha_1  A\nonumber\\&&\hspace{0.5cm}+\alpha_2 +2 \mathcal{B}_g)+2 B R^6 (\alpha_2 +2 \mathcal{B}_g)\big]
\end{eqnarray}

\section{Physical analysis of the Strange star models}\label{sec5}

\subsection{Regular Behavior of Minimally Deformed SS models} \label{sec5.1}
\begin{figure*}[!htb]
    \centering
    \includegraphics[width=8cm,height=5cm]{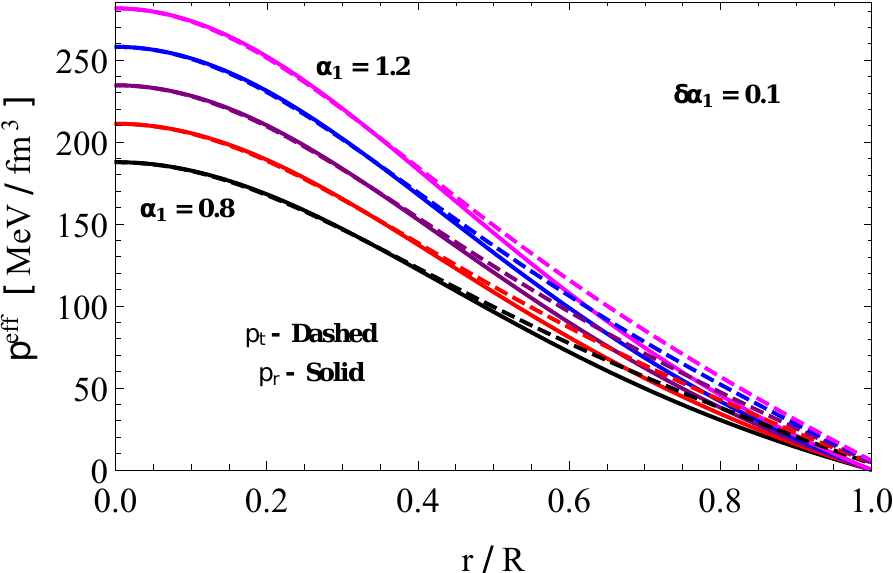}~~~ \includegraphics[width=8cm,height=5cm]{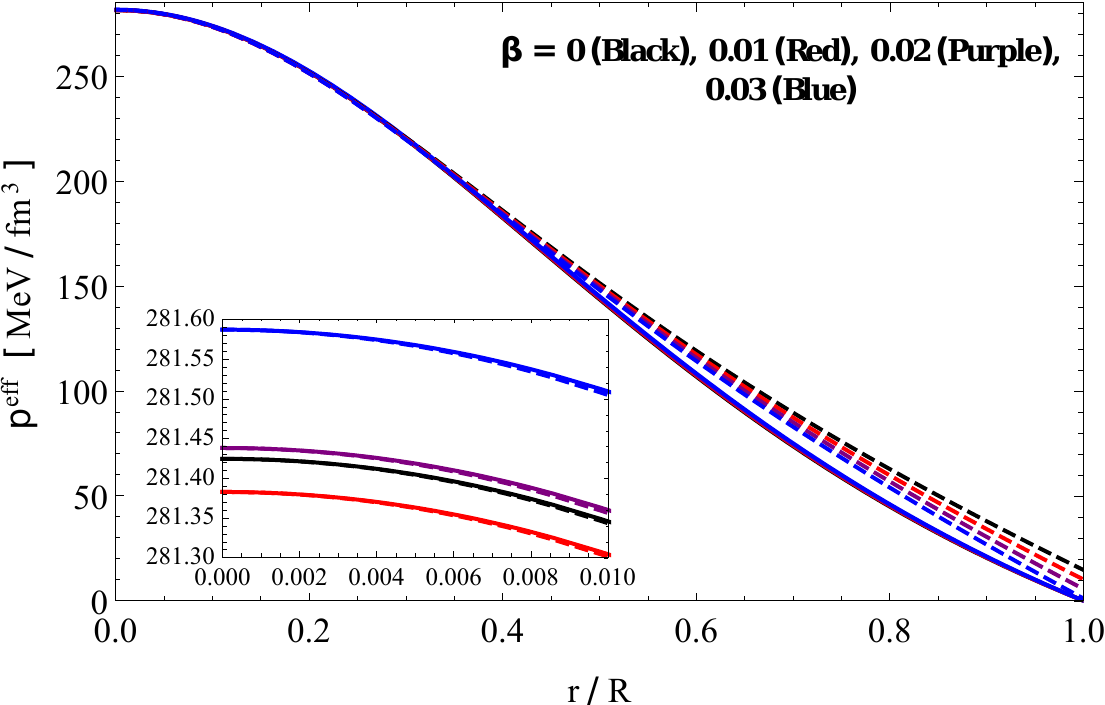}\\
     \includegraphics[width=8cm,height=5cm]{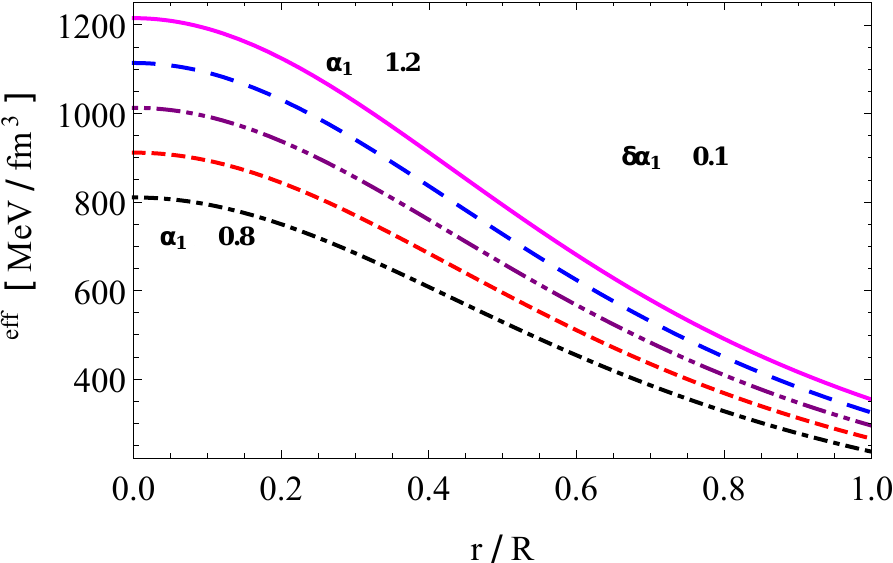}~~~  \includegraphics[width=8cm,height=5cm]{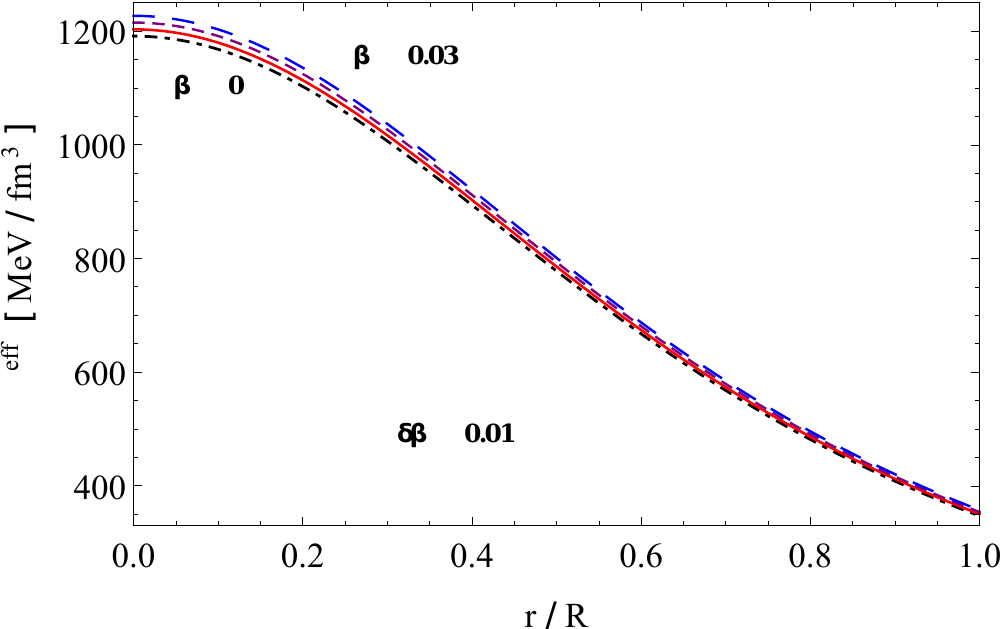}
    \caption{\textit{Top panels} and \textit{Bottom panels} show the  pressures [ radial ($p^{\text{eff}}_r$) and tangential ($p^{\text{eff}}_{t}$)] and energy density ($\rho^{\text{eff}}$) with respect to $r/R$ for different $\alpha_1$ and  $\beta$, respectively for the $\theta^0_0=\rho$ solution. 
 We set the numerical values $~A =0.011 /km^2,~B = 2.3\times 10^{-6}/km^4,~R =10.5 \,km$, and $\alpha_2=10^{-46}/km^2$ for plotting of left panels and right panel when $\beta= 0.02$ and $\alpha_1= 1.2\, km^2$, respectively.}
    \label{fig2}
\end{figure*}

In this section we provide a detailed physical analysis of our solutions based on the graphical plots presented here. Let us start by analysing the $\theta^0_0=\rho$ solution. The top left panel in Figure \ref{fig2} displays the effective radial pressure at each interior point of the stellar configuration. The pressure falls off smoothly as one moves from the center towards the boundary. At the boundary the effective radial pressure vanishes as we expect as there is no energy flux to the exterior spacetime. Contributions from the nonmetricity scalar, $\cal Q$, represented by $\alpha_1$ shows that the effective radial pressure increases as $\alpha_1$ increases. A similar observation applies to the effective tangential pressure. The top right panel shows the radial and transverse stresses when the decoupling parameter is increased. There are very small deviations in both the effective radial and transverse pressures. We also note that the effective tangential pressure dominates its radial counterpart in the surface layers of the compact object. In Figure \ref{fig2} (bottom left panel), we observe that the density is a monotonically decreasing function of the scaled radial coordinate, $\frac{r}{R}$. We have fixed the deformation parameter, $\beta$ and varied the $\cal Q$ switch represented by $\alpha_1$. It is clear that as $\alpha_1$ increases, the density of the compact object increases. In the bottom right panel of Figure \ref{fig2}, we kept $\alpha_1$ constant and varied the deformation parameter. In this scenario the effective density shows very little variation when $\beta$ increases. 

\begin{figure*}[!htb]
    \centering
   \includegraphics[width=8cm,height=5cm]{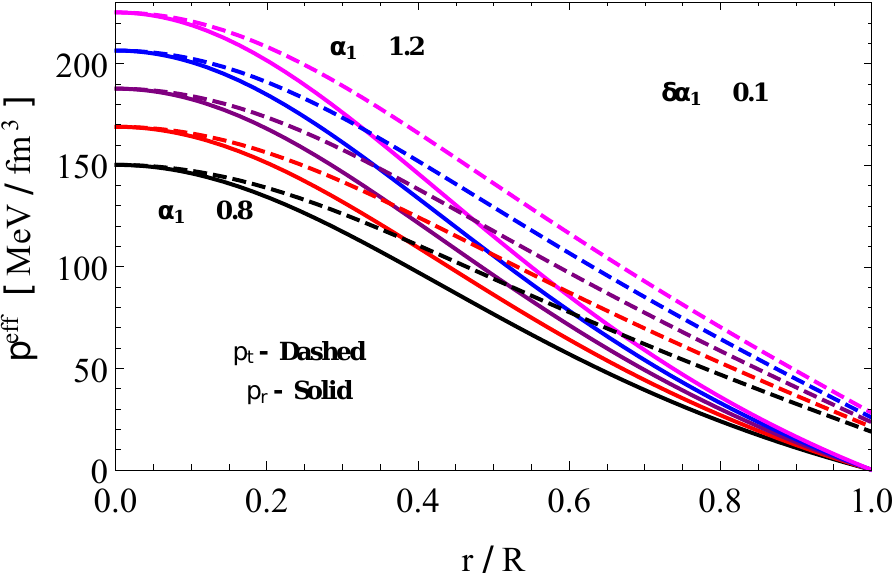}~~~ \includegraphics[width=8cm,height=5cm]{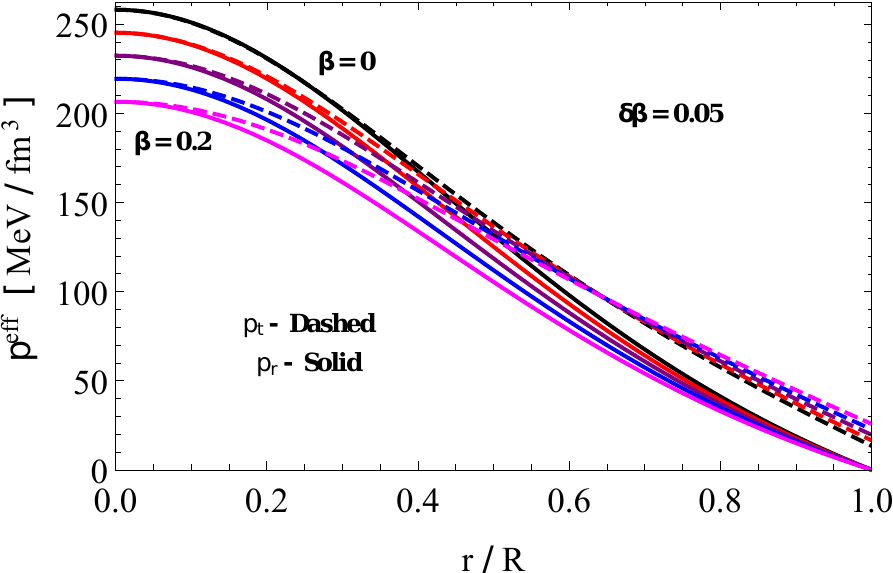}\\
     \includegraphics[width=8cm,height=5cm]{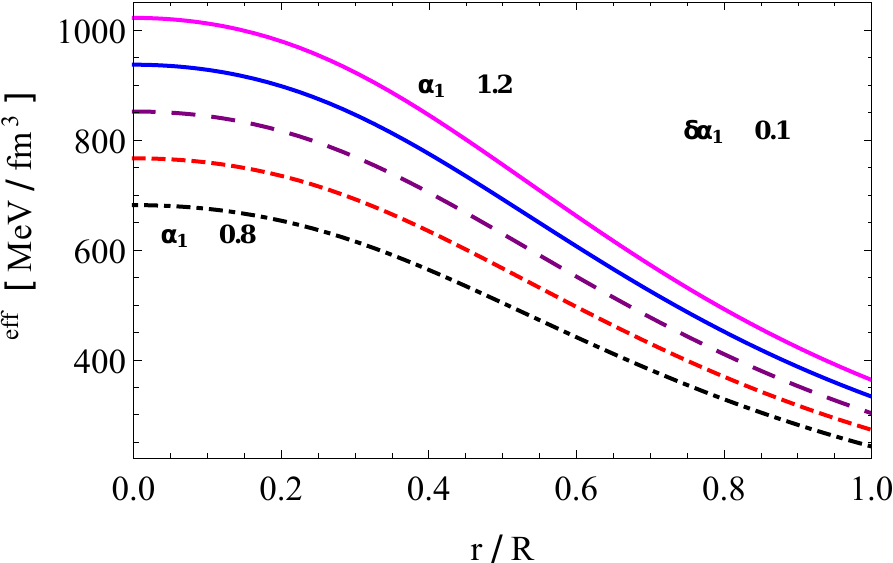}~~~  \includegraphics[width=8cm,height=5cm]{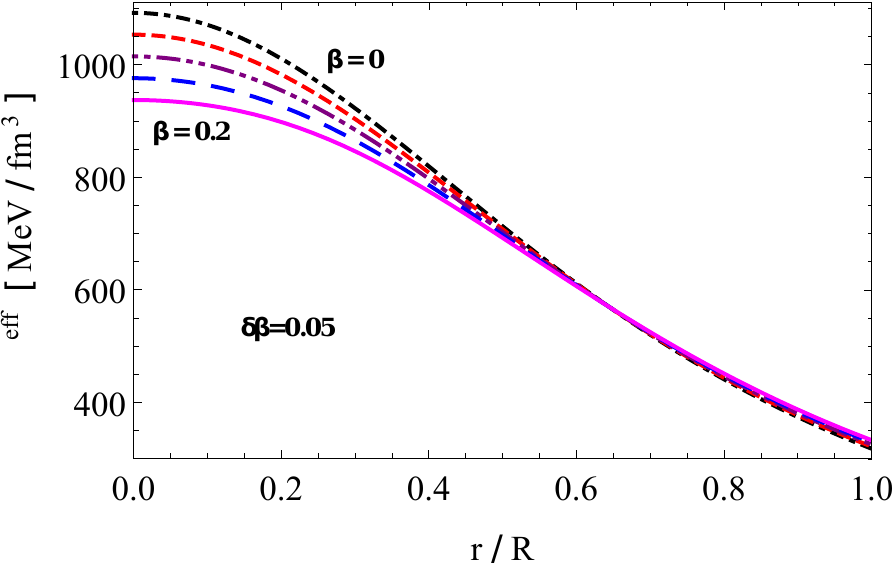}
    \caption{\textit{\textit{Top} and \textit{Bottom} panels show the pressures [radial ($p^{\text{eff}}_r$) and tangential ($p^{\text{eff}}_{t}$)] and energy density ($\rho^{\text{eff}}$) with respect to  $r/R$ for different $\alpha_1$ and  $\beta$, respectively for $\theta^1_1=p_r$ solution. 
 We set the numerical values $~A =0.011 /km^2,~B = 2.3\times 10^{-6}/km^4,~
R = 10.5\,km$, $\alpha_2=10^{-46}/km^2$ for plotting of left panels and right panel when $\beta= 0.2$ and $\alpha_1= 1.1\, km^2$, respectively.}}
    \label{fig3}
\end{figure*}

We now turn our attention to the $\theta^1_1=p_r$ solution. In Figure \ref{fig3}, we present the effective stresses and density when 
\begin{itemize}
    \item [(1)]$\alpha_1$ is varied and the deformation parameter, $\beta$ is fixed.
    \item[(2)] $\alpha_1$ is fixed and $\beta$ is varied.
    \end{itemize}
    The top right panel of Figure \ref{fig3} shows that the effective tangential pressure dominating the radial pressure as one approaches the boundary. As in the $\theta^0_0=\rho$ solution, an increase in $\alpha_1$ increases both the radial and tangential pressures throughout the configuration. When $\beta$ is varied and $\alpha_1$ is fixed (Figure \ref{fig3} top right panel) we note that the pressures are monotonically decreasing functions of the scaled radial coordinate. A peculiar observation is the switching of the effective tangential stress within the core. For some finite radius, $p^{\text{eff}}_{t}$ decreases as the decoupling parameter increases. This phenomenon has been observed in models of compact stars within the framework of Einstein-Gauss-Bonnet gravity. The bottom panels of Figure \ref{fig3} reveal the behaviour of the density profile. The behaviour of the density is very similar to the $\theta^0_0$-$\rho$ sector. The effect of $\beta$ is to suppress any increase in the density as one approaches the surface. This means that the decoupling parameter allows for higher densities in the central core regions of the star. 

\begin{figure*}
    \centering
    \includegraphics[width=8.5cm,height=5.5cm]{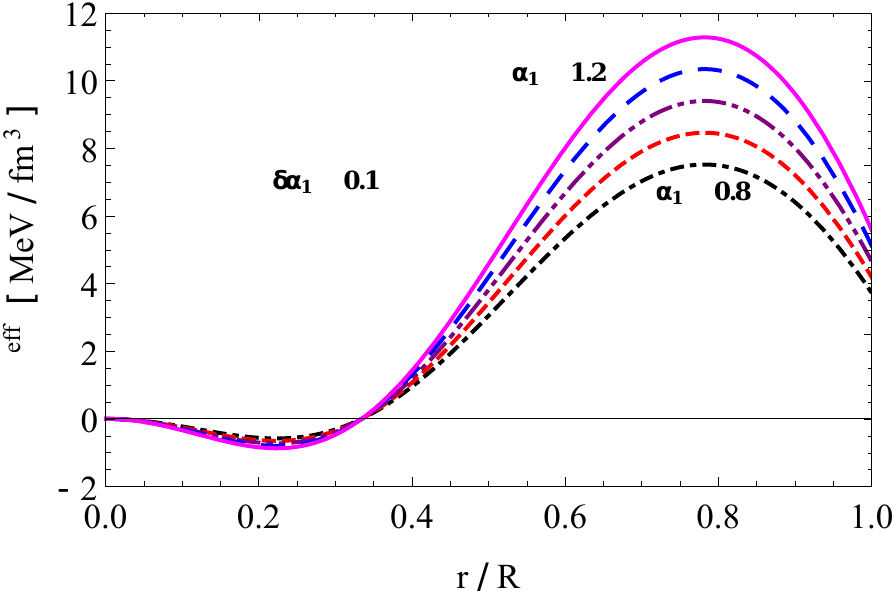} ~~~ \includegraphics[width=8.5cm,height=5.5cm]{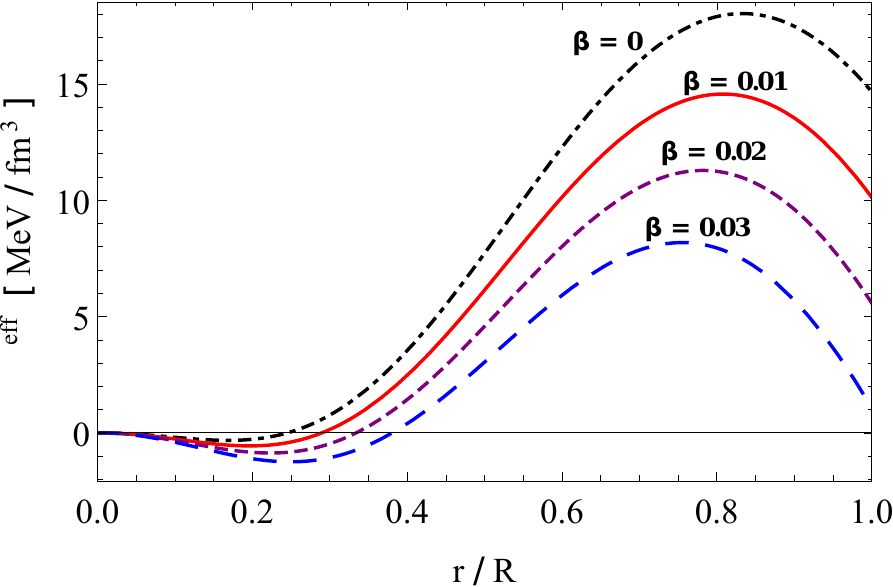}
     \includegraphics[width=8.5cm,height=5.5cm]{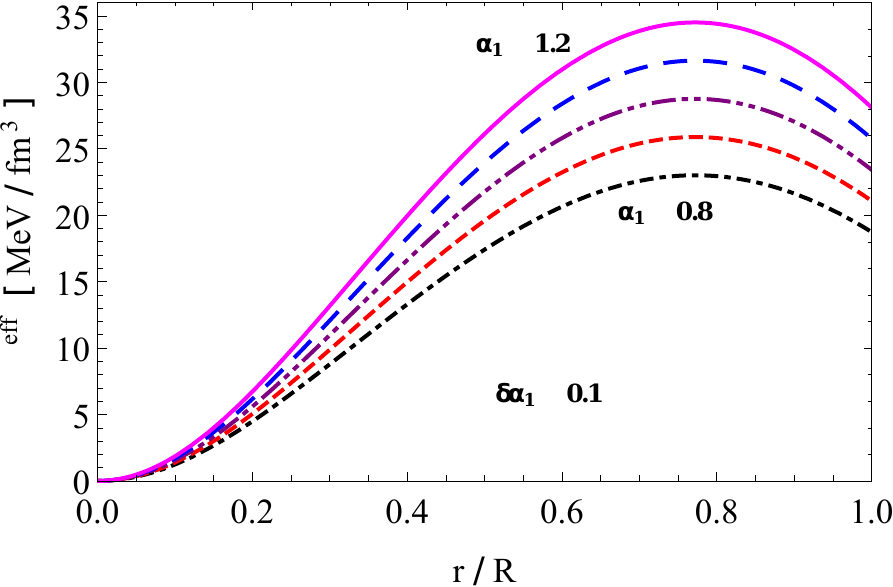} ~~~ \includegraphics[width=8.5cm,height=5.5cm]{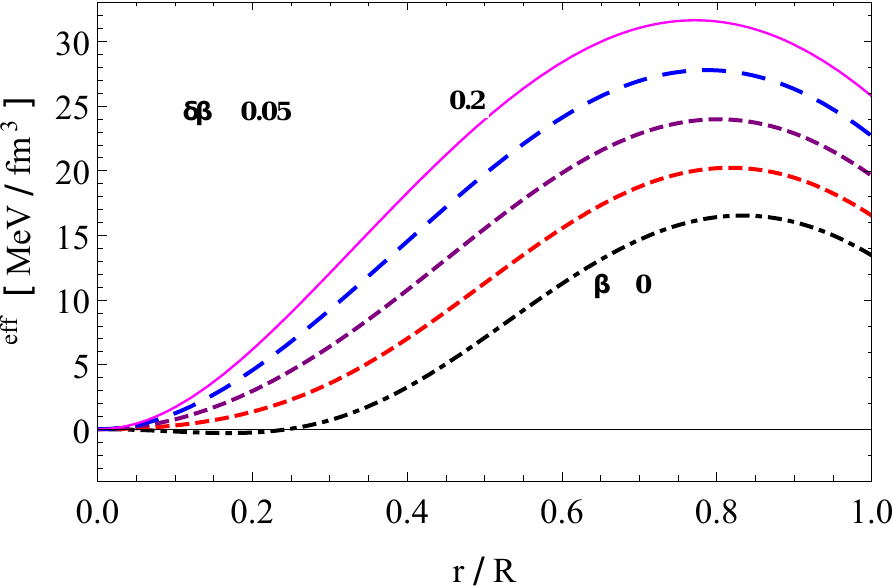}
    \caption{\textit{Top panels} (for solution $\theta^0_0=\rho$) and \textit{Bottom panels}  (for solution $\theta^1_1=p_r$) show the effective anisotropy ($\Pi^{\text{eff}}$) with respect to $r/R$ for different $\alpha_1$ and  $\beta$. We set same numerical values as used in Figure \ref{fig2} and  Figure \ref{fig3}.}
    \label{fig4}
\end{figure*}

    We have plotted the effective anisotropy parameter in Figure \ref{fig4}. The top left panel shows the trend in $\Pi^{\text{eff}}$ when the $\cal Q$ switch is varied and the decoupling constant 
    is held fixed. We observe that the anisotropy parameter is negative from the center up to a certain radius, $r_0$. This implies that the effective radial pressure dominates 
    it tangential counterpart giving rise to an attractive force due to anisotropy. As one moves away from $r = r_0$ towards the boundary, anisotropy becomes positive, signifying a repulsive force which stabilizes the surface layers of the star. In the top right panel we observe the behaviour of $\Pi^{\text{eff}}$ when the decoupling constant is varied. We see that the anisotropy parameter decreases as $\beta$ increases suggesting that the decoupling constant has a quenching effect on the contributions due to anisotropy. Here too, anisotropy is negative in the central regions of the star, indicating unstable regions here compared to the repulsive contribution from $\Pi^{\text{eff}}$ at the surface. For the $\theta^1_1=p_r$ solution, we have plotted the effective anisotropy in the bottom panels of Figure \ref{fig4}. In the left panel we note that the anisotropy parameter is positive throughout the stellar configuration. This gives rise to a repulsive force which helps counteract the inwardly driven gravitational force. In the right panel, it is interesting to observe that an increase in $\beta$ leads to an increase in the anisotropy thus strengthening the force due to pressure anisotropy. 

    \subsection{Stability Analysis of SS Models via Harrison-Zeldovich-Novikov (HZN) criterion} \label{sec5.2}
\begin{figure*}[!htb]
    \centering
    \includegraphics[width=8cm,height=5cm]{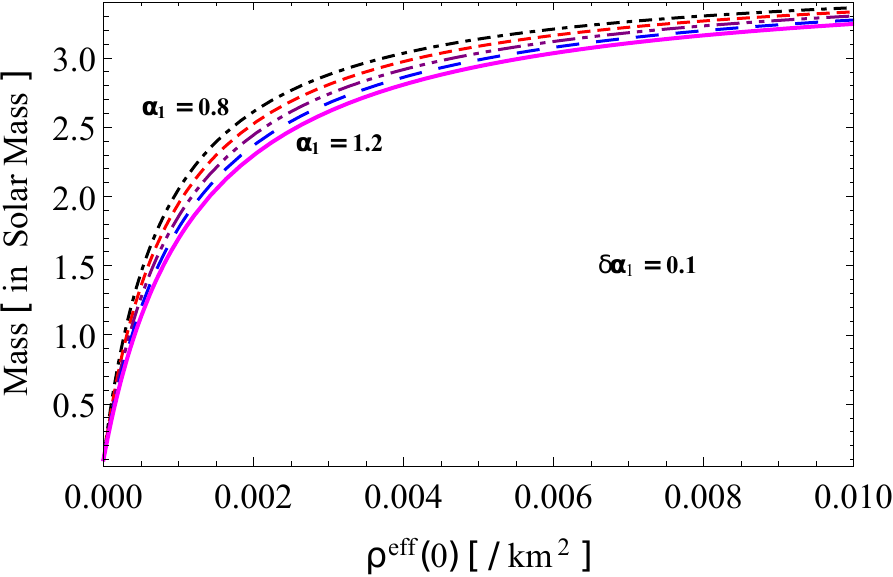}~~~ \includegraphics[width=8cm,height=5cm]{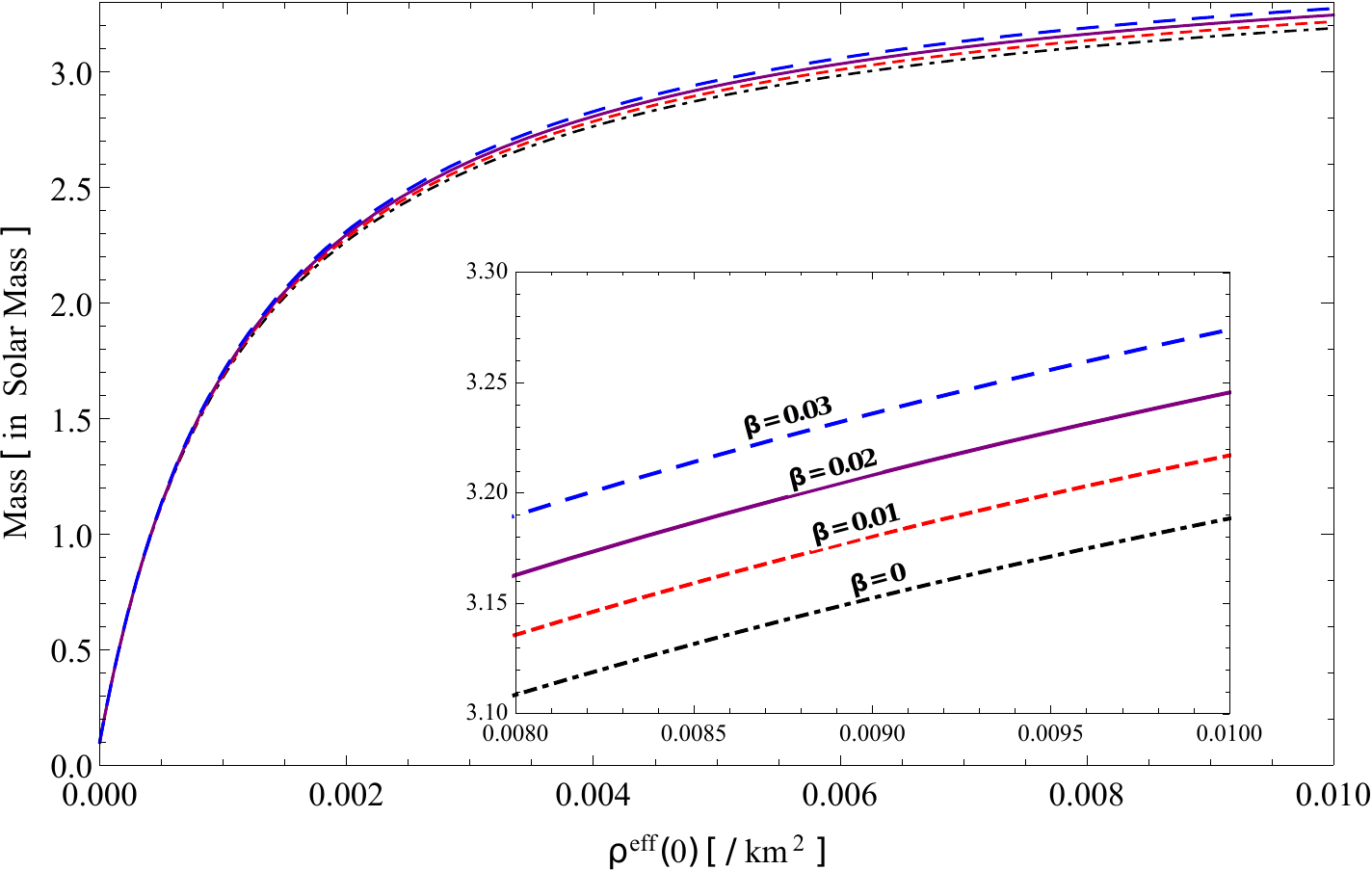}\\
    \includegraphics[width=8cm,height=5cm]{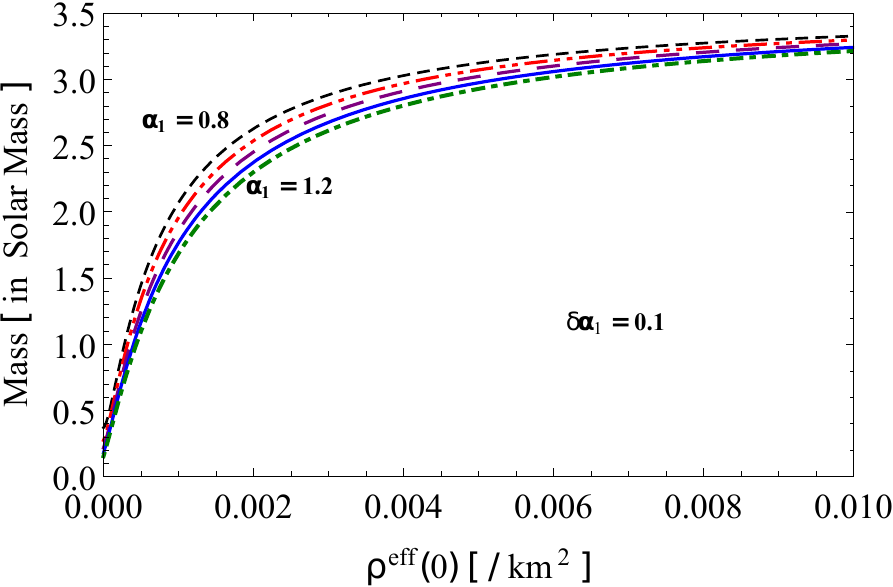}~~~ \includegraphics[width=8cm,height=5cm]{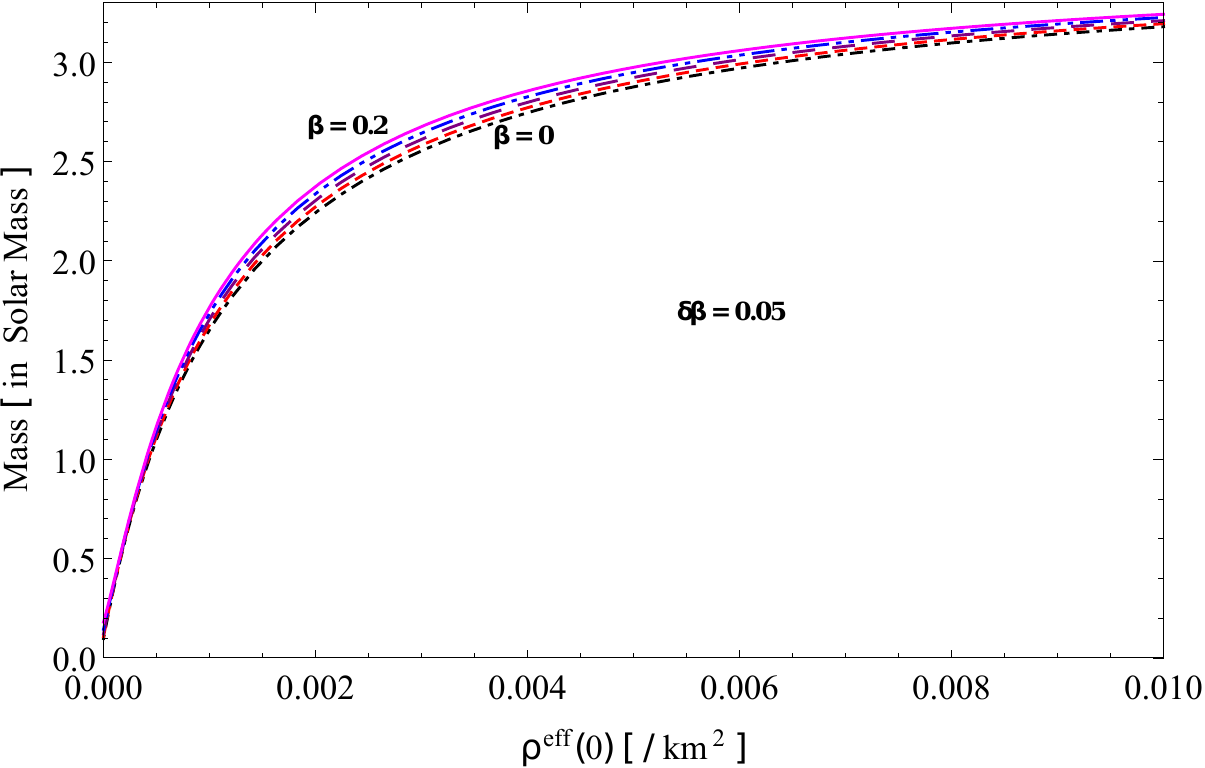}
    \caption{Mass versus central density for different $\alpha_1$ and  $\beta$ for the solution $\theta_0^0=\rho$--top panels and $\theta_1^1=p_r$--bottom panels. We set same numerical values as used in Figure \ref{fig2} and  Figure \ref{fig3}.}
    \label{fig5}
\end{figure*}

    In Figure \ref{fig5}, we subject our solutions to the Harrison-Zeldovich-Novikov (HZN) stability criterion. The HZN stability criterion \cite{harris,zeld} hinges on the following constraints:
    \begin{eqnarray}
    \frac{dM}{d\rho_c} &>& 0 \hspace{0.5cm} \rightarrow \mbox{stable configuration}\\ \nonumber \\
  \frac{dM}{d\rho_c} &<& 0 \hspace{0.5cm} \rightarrow \mbox{unstable configuration}  
        \end{eqnarray}
  In Figure \ref{fig5}, we observe that $\frac{dM}{d\rho_c} > 0$ for all our models thus indicating that these configurations are stable. The top and bottom left panels show that an increase in the $\cal Q$ parameter leads to an {\em increase} in the stability of the bounded configurations, albeit that the increase in mass as a function of the effective central density is relatively small. Similar observations are true when the decoupling constant is increased. It is interesting to observe that the M-$\rho(0)$ plots in Figure \ref{fig5} display very similar trends in behaviour and magnitude, whether the $\cal Q$ parameter is increased and $\beta$ is held constant or vice versa.
\begin{figure*}[!htb]
    \centering
    \includegraphics[width=8cm,height=5.8cm]{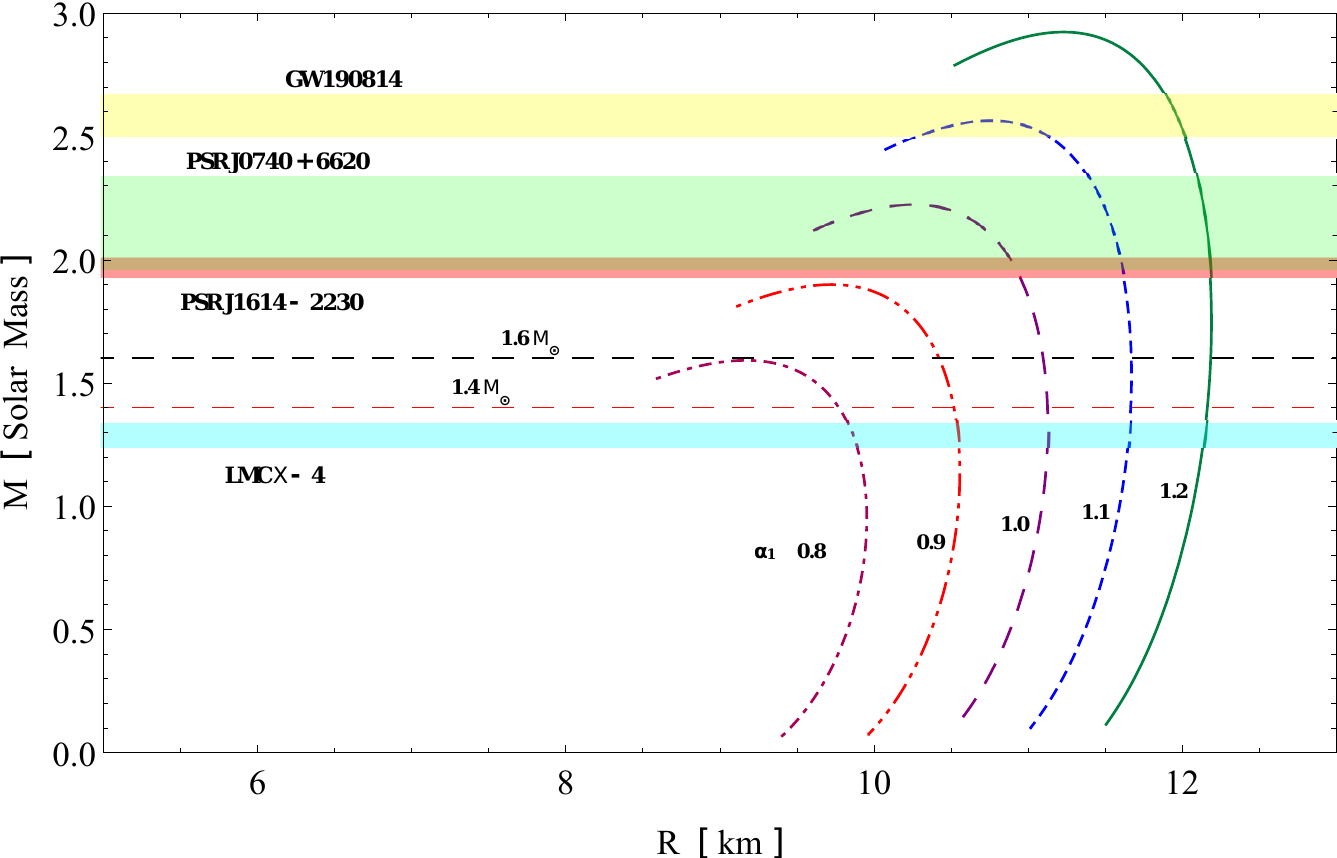}~~~~\includegraphics[width=8.1cm,height=5.7cm]{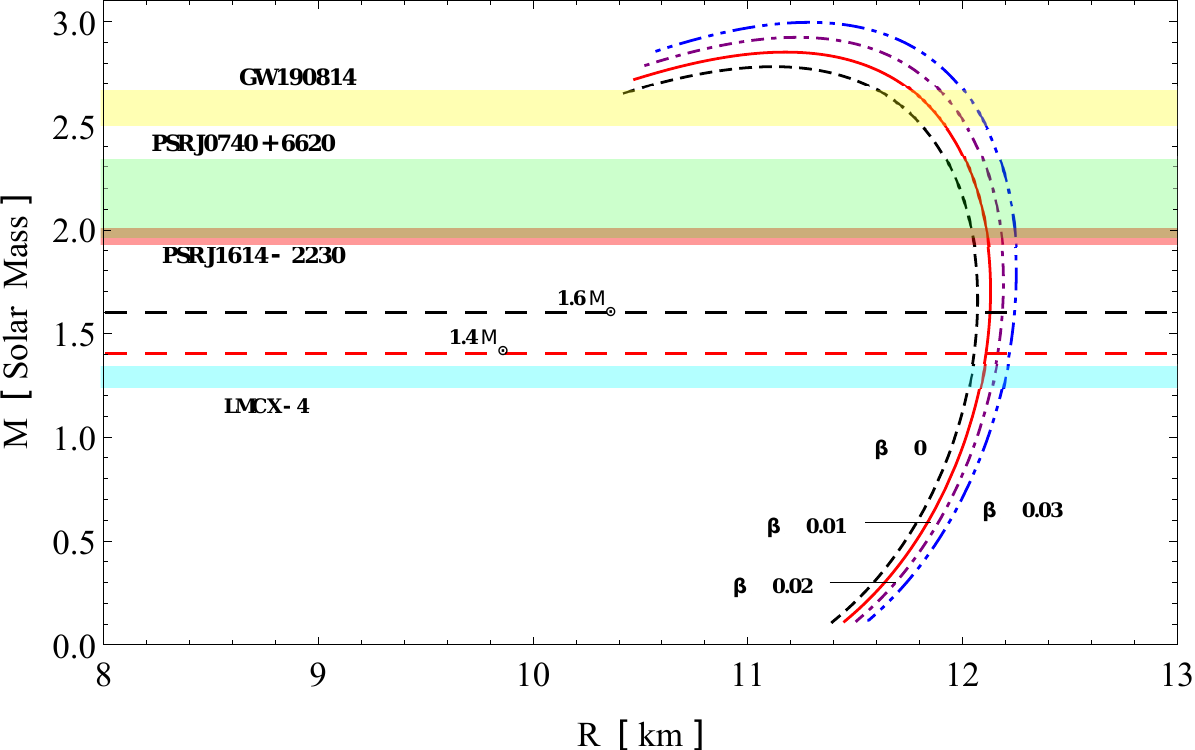}
    \caption{The left and right panels show the $M-R$ curves depending on different values $\alpha_1$ and  $\beta$, respectively when $\theta^0_0=\rho$.}
    \label{fig6}
\end{figure*}
\begin{figure*}[!htb]
    \centering
    \includegraphics[width=8cm,height=5.8cm]{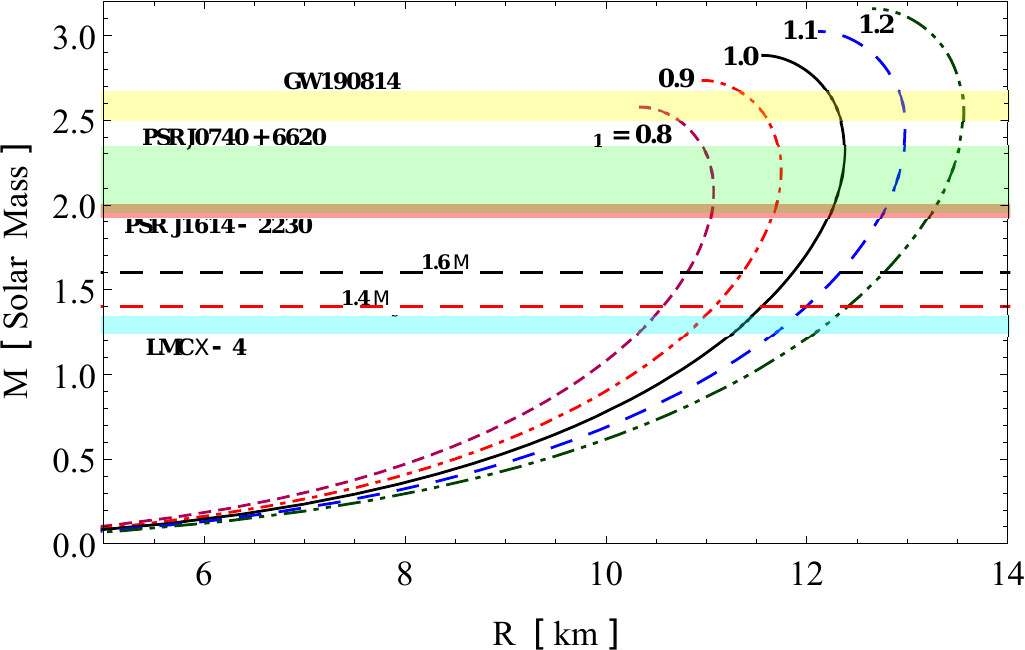}~~~ \includegraphics[width=8cm,height=5.8cm]{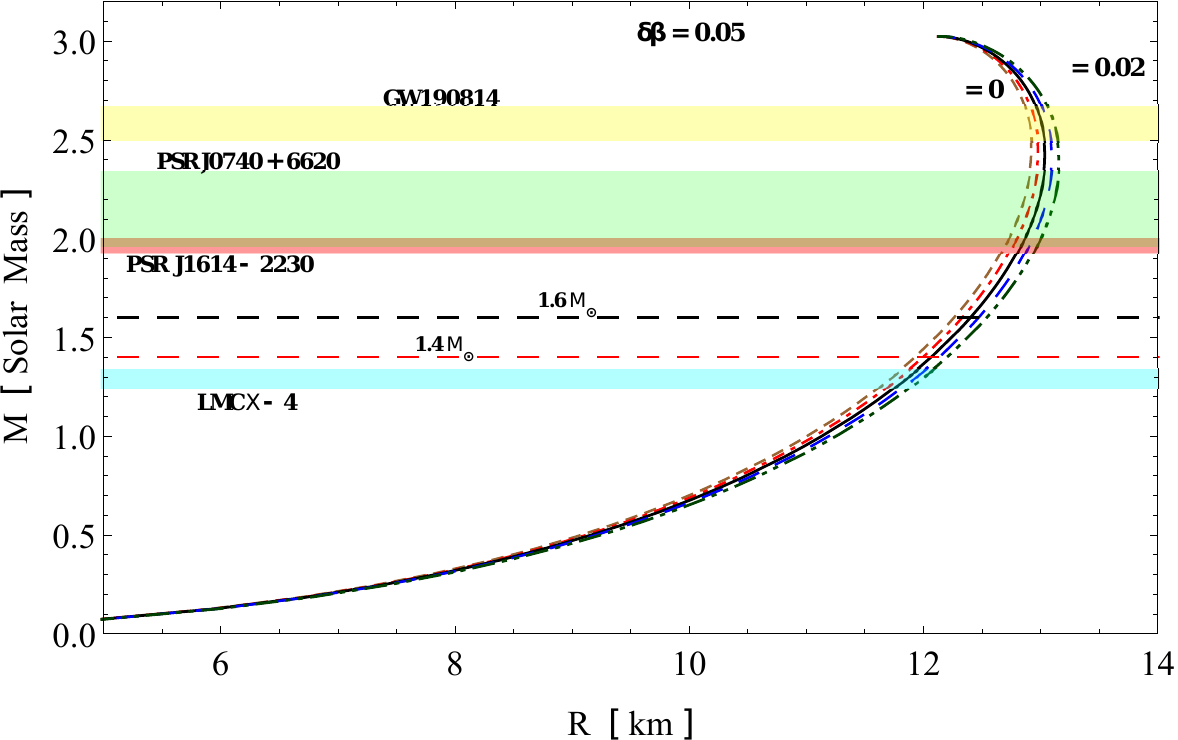}
    \caption{The left and right panels show the $M-R$ curves depending on different values of $\alpha_1$ and  $\beta$, respectively when $\theta^1_1=p_r$. }
    \label{fig7}
\end{figure*} 

  \subsection{Stability Analysis via Adiabatic Index} \label{sec5.3}
We now focus on the stability of our models by employing the adiabatic stability criterion which was first derived by Chandrasekhar \cite{chandra1,chandra2}  for isotropic pressure profiles. The adiabatic stability criterion is given by \begin{equation}
 \Gamma = \left(1+\frac{\rho}{p}\right) \left(\frac{dp}{d\rho}\right)_S\end{equation} with the limiting case of $\Gamma > 4/3$ for bounded configurations having isotropic pressure, $p$, and $\frac{dp}{d\rho}$ is the velocity of sound. The subscript $S$ denotes a constant specific entropy. Herrera and co-workers \cite{Herrera1,Herrera2} demonstrated that this condition gets modified in the presence of anisotropy and dissipation.  The stability criterion is modified in the presence of pressure anisotropy and assumes the form
\begin{equation} \label{modgamma}
\Gamma < \frac{4}{3} + \left[-\frac{4}{3}\frac{(p^{\text{eff}}_r - p^{\text{eff}}_t)}{|(p^{\text{eff}}_r)'|r}\right]
\end{equation}
where the prime denotes differentiation with respect to $r$.  The second term in (\ref{modgamma}) arises from relativistic contributions and the vanishing of this term yields the Newtonian limit, $\Gamma < \frac{4}{3}$ for {\em unstable} regions. Dissipation in the form of a radial heat flux or the presence of density inhomogeneities can alter the adiabatic index. The critical value for the adiabatic index, ($\Gamma_{crit}$) is defined as  \cite{mou}
\begin{equation}
\Gamma_{crit} = \frac{4}{3} + \frac{19}{21}u\end{equation}
where  $(u = M/R)$ is the compactness of the stellar model. Stability against radial perturbations is ensured when $\Gamma > \Gamma_{crit}$ \cite{mou}. It has been recently shown that stable neutron star configurations, together with white dwarfs and more massive compact objects are achieved for $\Gamma$ ranging between 2 and 4 \cite{ayan3}. 
Table \ref{table1} displays the stability criterion values obtained for our models. For the  $\theta^1_1=p_r$ models we observe that $\Gamma = 1.72$ and is independent of the variation in $\alpha_1$.  We also note that $\Gamma_{crit}$ is independent of contributions from  $\cal Q$. On the other hand, both $\Gamma$ and $\Gamma_{crit}$ increase as the decoupling parameter increases thus indicating that the compact objects become more stable as $\beta$ is increased. This increase in stability has also been observed EGB stars in the presence of higher values of $\beta$. 
 \begin{table*}[!htb]
\centering
\caption{The calculated values of the adiabatic index $\Gamma$ and its critical value $\Gamma_{crit}$ for different values of $\alpha_1$ and $\beta$. }\label{table1}
 \scalebox{0.96}{\begin{tabular}{| *{12}{c|} }
\hline
\multirow{2}{*}{Solution} & & \multicolumn{5}{c|}{$\alpha_1$} & \multicolumn{4}{c|}{$\beta$} & \\
\cline{2-12}
&  & 0.8 & 0.9 & 1.0 & 1.1 & 1.2 & 0 & 0.01 & 0.02 & 0.03 &   \\ 
\hline
\multirow{2}{*}{$\theta^0_0(r)=\rho(r)$ } &  $\Gamma$  & 1.72  & 1.72 & 1.72 &  1.72 & 1.72 & 1.744 & 1.733 & 1.722 & 1.711 & \\
\cline{2-12}
  &  $\Gamma_{crit}$  & 1.506  & 1.506 & 1.506 & 1.506 & 1.506 & 1.503 & 1.505 & 1.506 & 1.508 &  \\
\hline
\hline 
\multirow{2}{*}{Solution} & & \multicolumn{5}{c|}{$\alpha_1$} & \multicolumn{5}{c|}{$\beta$} \\
\cline{2-12}
&  & 0.8 & 0.9 & 1.0 & 1.1 & 1.2 & 0 & 0.05 & 0.1 & 0.15 & 0.2 \\
\hline
\multirow{2}{*}{$\theta^1_1(r)=p_r(r)$} & $\Gamma$  &  3.52 & 3.52 & 3.52 & 3.52 & 3.52 & 1.743 & 1.958 & 2.266 & 2.739 & 3.519 \\
\cline{2-12}
 &  $\Gamma_{crit}$ & 1.503 & 1.503   & 1.503  & 1.503 & 1.503 & 1.503 & 1.503 & 1.503 & 1.503 & 1.503   \\
\hline
\end{tabular}}
\end{table*}
\subsection{Measurements of the Constraint on Maximum Mass Limit of Strange Stars via $M-R$ Diagrams} \label{sec5.4}  
The mass profiles as a function of radii are plotted in Figure \ref{fig6} for the $\theta^0_0=\rho$ sector. It is clear from both the left and right panels that these models can account for the existence of compact objects with masses between $1.29M_\odot$ and $2.8M_\odot$. This is an interesting observation as the predicted masses are above the accepted values for neutron stars ($\approx 2.0M_\odot$). The upper values for the mass profiles adequately account for the secondary component of the GW190814 event which in the current literature is purported to be a remnant with mass range of $2.6M_\odot$. It is thought that this remnant is a product of a bNS merger with a typical mass in the range of  $2.5M_\odot$ to  $2.9M_\odot$ or even as high as $3.4M_\odot$. In Figure \ref{fig7},  we have displayed the $M-R$ plots for the $\theta^1_1=p_r$ sector. Then it reveals that an increase in the $\cal Q$ parameter results in a significant increase in mass of the stellar configuration. We also note that the upper mass limit for varying $\cal Q$ parameter is greater than $3M_\odot$. In addition, the predicted radii for higher mass configurations are larger than their  $\theta^0_0=\rho$ sector counterparts. The right panel of Figure \ref{fig7} reveals no large deviations in the mass profiles as the decoupling parameter is increased. Any interesting observation is that predicted masses for known compact objects such as LMC X-4, PSR J1614+2230 and PSR J0740+6620 occur for larger radii compared to models obtained when the $\cal Q$ parameter is varied, while $\beta$ is held fixed. In essence, we are able to predict stellar models with larger radii and masses when $\beta$ is varied. We present the predicted radii of well-known compact objects in Tables \ref{table2} ($\theta^0_0=\rho$ sector) and \ref{table3} ($\theta^1_1=p_r$ sector). In Table \ref{table2}, the variation of $\cal Q$ parameter embodied in $\alpha_1$ (with $\beta$ held constant) yield radii with a upper value of $12.18$ km, while a varying $\beta$ gives an upper radius of $12.24$ km. Table \ref{table3} reveals that an upper bound on the radius of familiar compact objects is approximately 13.40km. If we extrapolate these predictions to the secondary component of the GW190814 event we obtain a constraint of $10.42~\text{km} \leq R \leq 13.55~\text{km}$ when $\alpha_1$ is varied and the decoupling parameter is held constant. On the other hand, the bound on the radii when $\beta$ is varied and $\alpha_1$ is fixed gives $12.90~ \text{km} \leq R \leq 13.11~ \text{km}$. The detection of gravitational waves and it electromagnetic components from the GW170817 event has led researchers to theorize about the progenitor leading to the $1.5M_\odot$ remnant. Several theories have been put forward which tend to rule out small radii of compact stars, very soft of very stiff equations of state. By appealing to microscopic nucleonic equations of state, Burgio et al. \citep{burgio} have shown that progenitor leading to the $R_{1.5}$ remnant of the GW170817 event must have had a radius in the range of $11.8~\text{km} \leq R \leq 13.1~\text{km}$. For the GW190814 event with a remnant mass of $2.6M_\odot$ our models predict radii in the range of $10.42 \leq R \leq 13.11~ \text{km} $ with a mass range of $2.5~ M_\odot$ to  $2.67~ M_\odot$. While the technique adopted for determination of stellar radii via the X-ray spectrum of compact objects residing in LMXBs reveal radii of in the range of $9.9-11.2~ \text{km}$ for $1.4-1.5~ M_\odot$, there have been counter arguments for larger radii.  \cite{lattimer} have argued that helium-containing atmospheres may account for larger stellar radii. Burgio et al. \citep{burgio} further argue that the source of GW170817 is a mixed binary system comprising Hybrid star and a Quark star. 
 \subsection{Measurements of the Constraint on Maximum mass Limit and the Range of Bag Constant of the Strange Stars via Equi-Plane Diagrams} \label{sec5.5} 
\begin{figure*}[!htb]
    \centering
    \includegraphics[width=8.5cm,height=6.5cm]{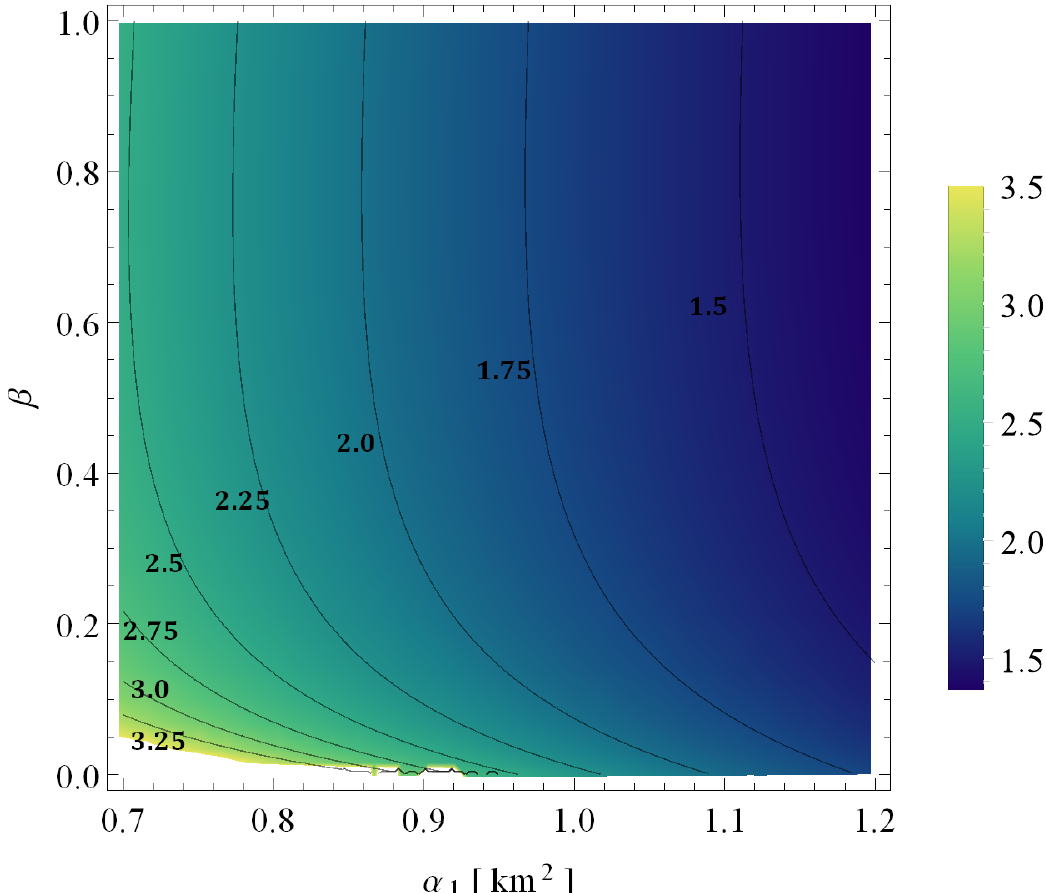}~~~~
     \includegraphics[width=8.5cm,height=6.5cm]{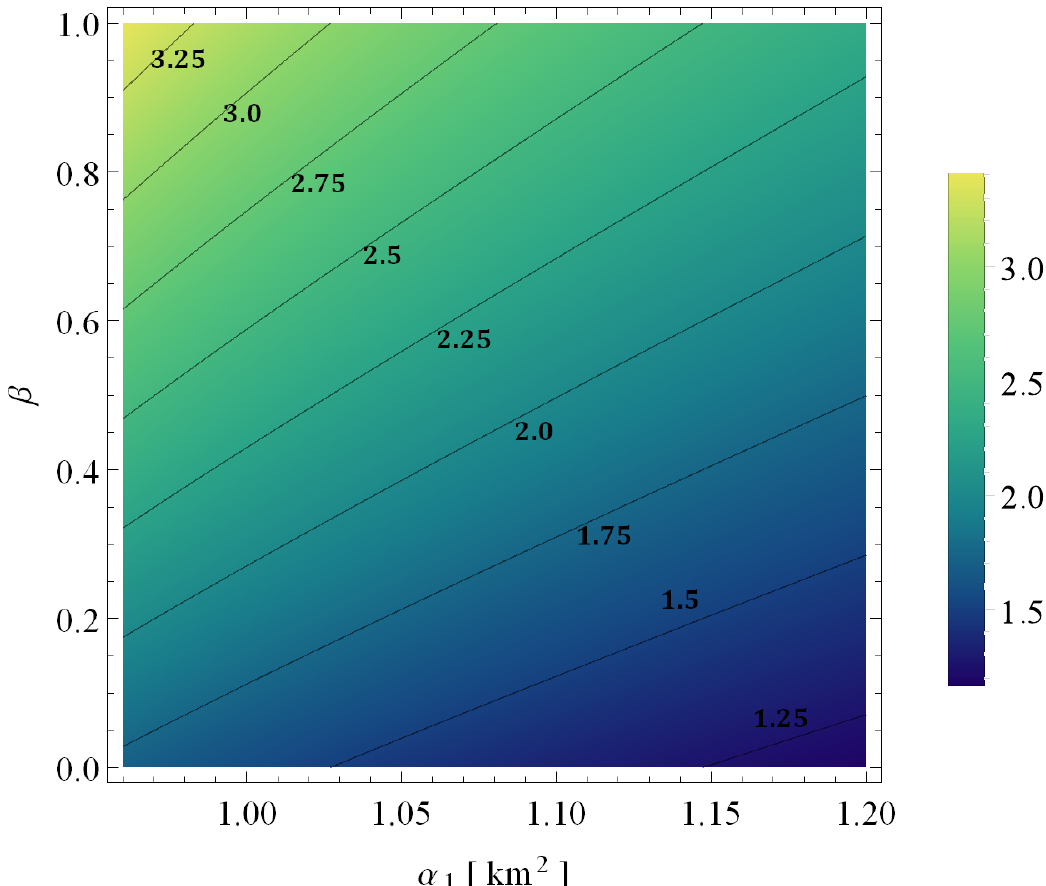}
\caption{\textit{Left panel:} $\alpha_1-\beta$ plane for equi-mass with $R =10.5 ~km,~ B=2.3\times 10^{-6}  /km^4~\mathcal{B}_g= 65 \,MeV/fm^3$ for the case $\theta^0_0=\rho$. \textit{Right panel:} $\alpha_1-\beta$ plane for equi-mass with $R =10.5 ~km,~ B=2.3\times 10^{-6}  /km^4~\mathcal{B}_g=65 \,MeV/fm^3$  for the case $\theta^1_1=p_r$.}
    \label{fig8}
\end{figure*}

\begin{figure*}[!htb]
    \centering
    \includegraphics[width=8.5cm,height=6.5cm]{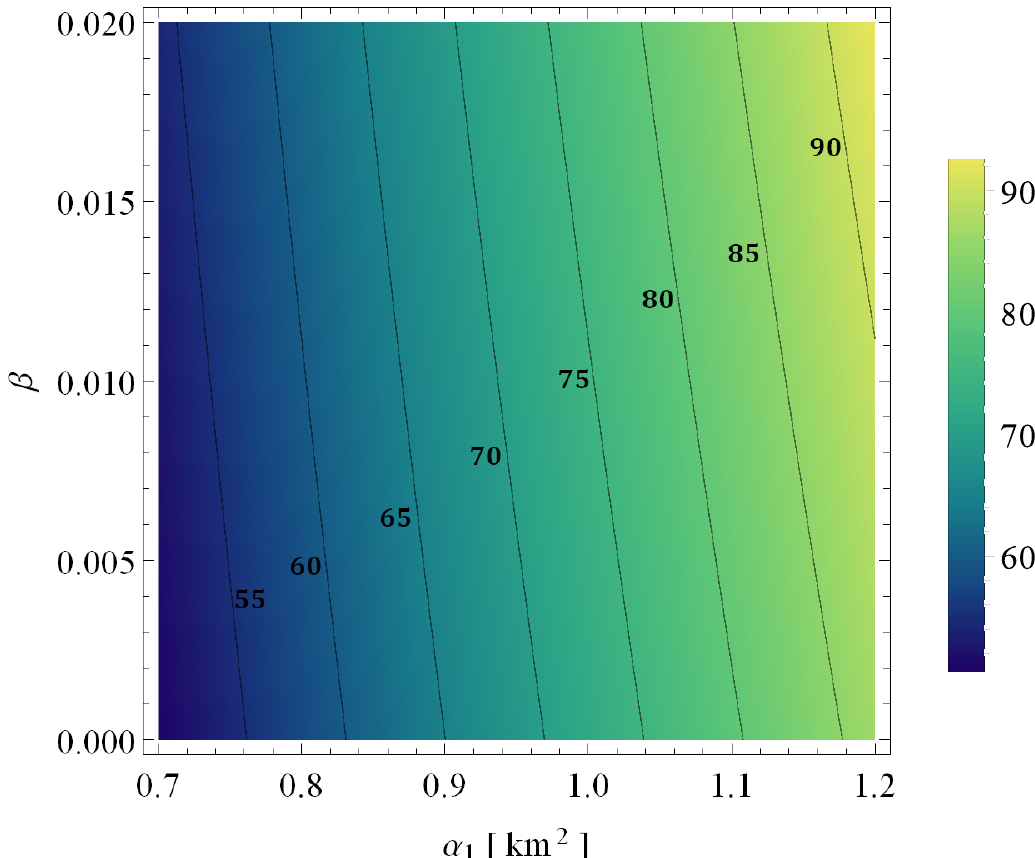}~~~~~~ \includegraphics[width=8.5cm,height=6.5cm]{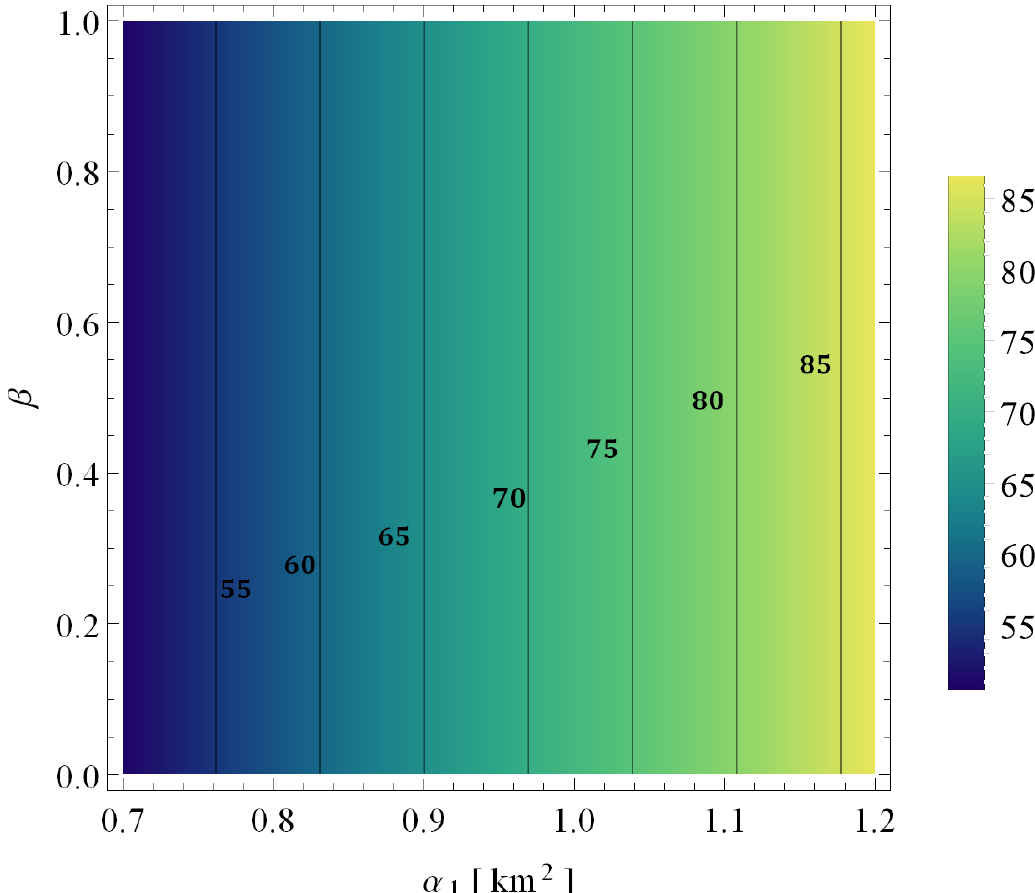} 
    \caption{\textit{Left panel}: $\alpha_1-\beta$ plane for equi-$\mathcal{B}_g$ with $R = 10.5~km,~B=2.3\times 10^{-6}/km^4$ for the case $\theta^0_0=\rho$.  \textit{Right panel}: $\alpha_1-\beta$ plane for equi-$\mathcal{B}_g$ with $R =10.5 ~km, ~B=2.3\times 10^{-6} /km^4$ for the case $\theta^1_1=p_r$. } 
    \label{fig9}
\end{figure*}

\begin{figure*}[!htb]
    \centering
   \includegraphics[width=8.5cm,height=6.5cm]{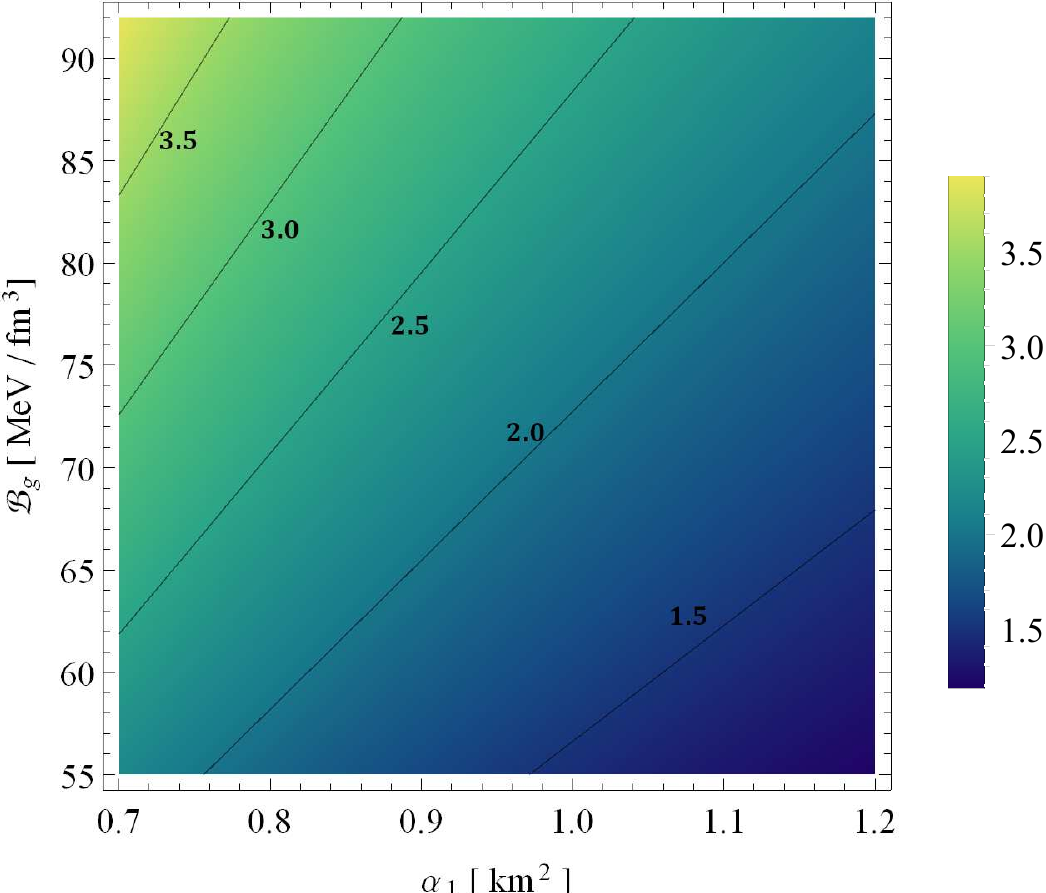}~~~~~~ \includegraphics[width=8.5cm,height=6.5cm]{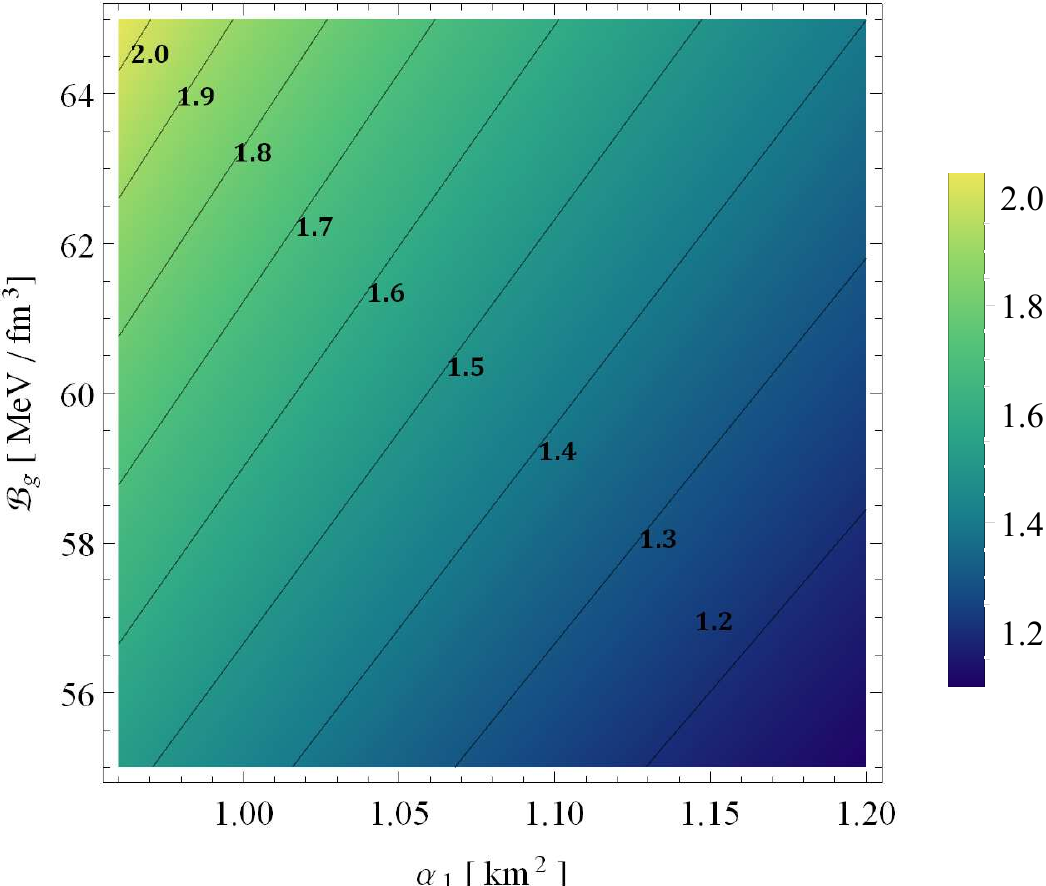}
\caption{\textit{Left panel}:  $\mathcal{B}_g-\alpha_1$ plane for equi-mass with $R =10.5 ~km, ~\beta =0.3,~B=2.3\times 10^{-6} /km^4$ for the case $\theta^0_0=\rho$.   \textit{Right panel}: $\mathcal{B}_g-\alpha_1$ plane for equi-mass with $R =10.5 ~km,~B= 2.3\times 10^{-6}/km^4,~\beta= 0.2$ for the case $\theta^1_1=p_r$.}
    \label{fig10}
\end{figure*}

\begin{figure*}[!htb]
    \centering
    \includegraphics[width=8.5cm,height=6.5cm]{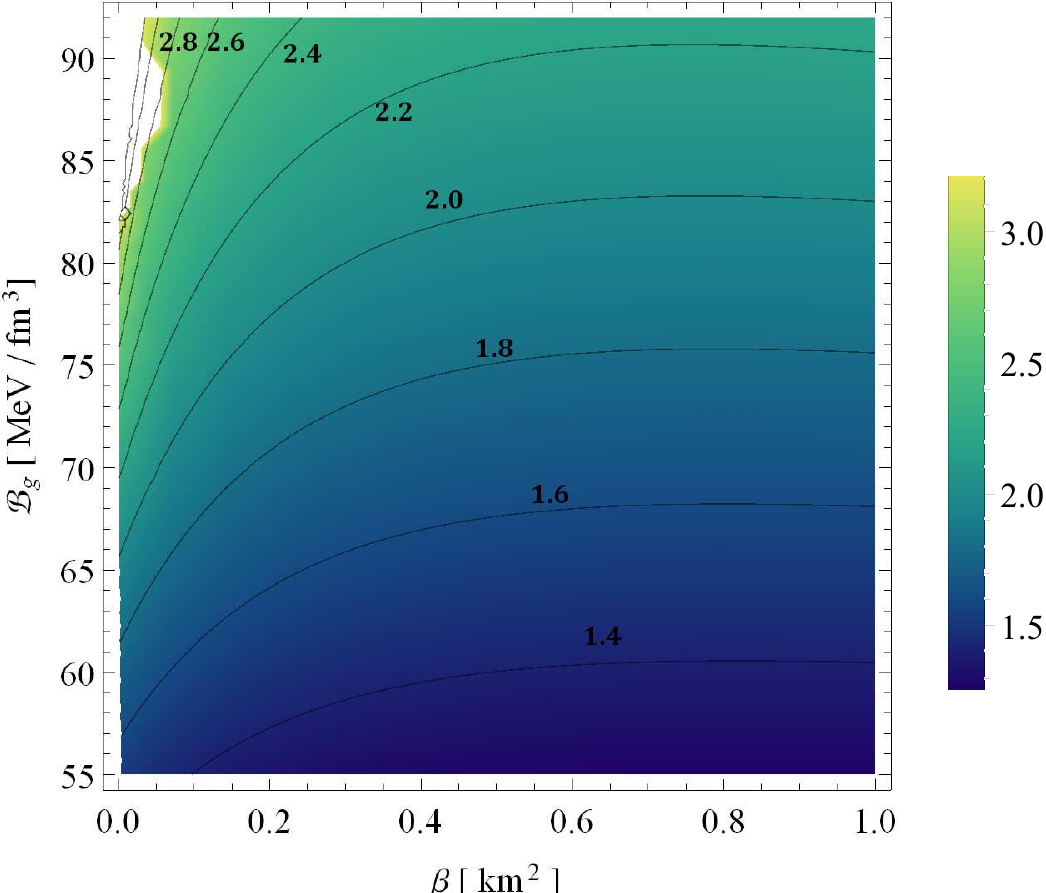}~~~\includegraphics[width=8.5cm,height=6.5cm]{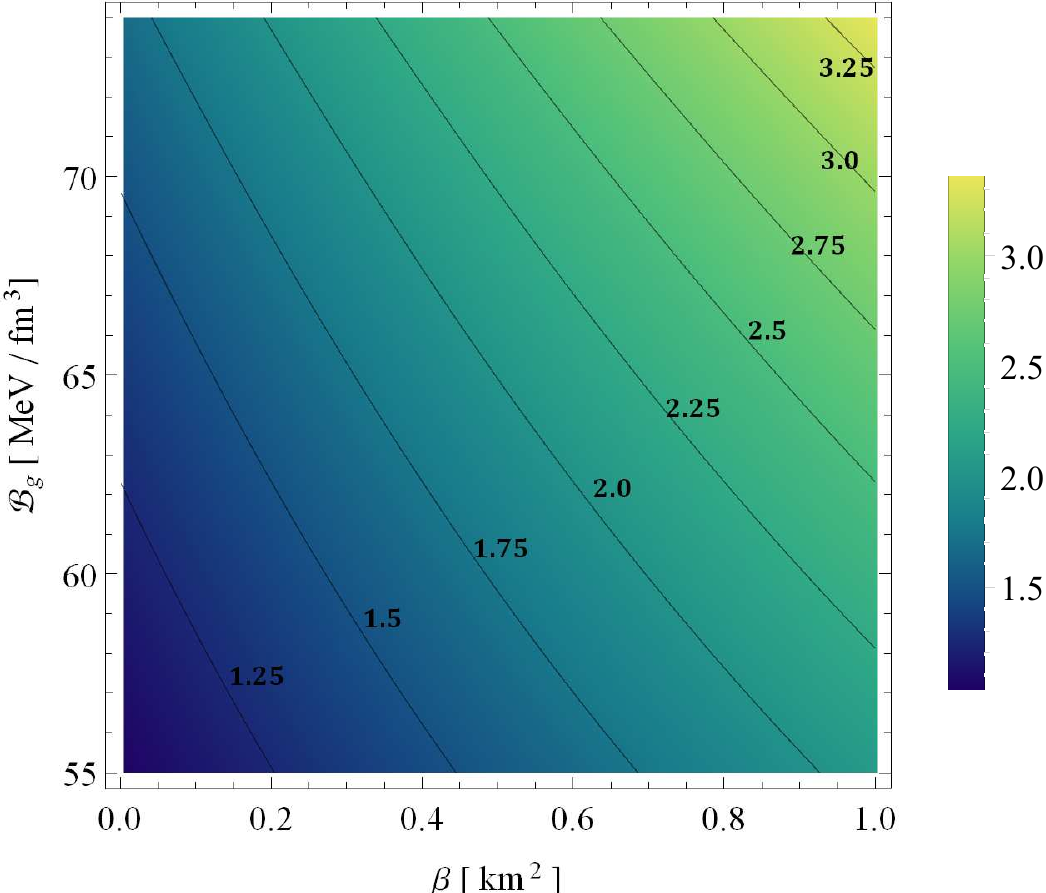}
    \caption{\textit{Left panels}:  $\mathcal{B}_g-\beta$ plane for equi-mass with $R=10.5 km,~\alpha_1= 1.1km^2,~B=2.3\times 10^{-6} /km^4$ for the case $\theta^0_0=\rho$.  \textit{Right panels}:  $\mathcal{B}_g-\beta$ plane for equi-mass with $R=10.5km,~\alpha_1= 1.1km^2,~B=2.3\times 10^{-6} /km^4$ for the case $\theta^1_1=p_r$.} 
    \label{fig11}
\end{figure*}

 Now we focus on the analysis of the Figs. \ref{fig8} - \ref{fig11}. The Figure \ref{fig8} is plotted for the $\alpha_1-\beta$ planes which show the equi-mass contours for a fixed radius $10.5 ~ \text{km}$ and bag constant $\mathcal{B}=65~ MeV/fm^3$. The left panel of Figure \ref{fig8} is plotted for  for the solution \ref{sec3.1} which is $\theta_0^0=\rho$ while the right panel is for the solution \ref{sec3.2} corresponding to $\theta_1^1=p_r$. 
It is observed from left figure that when we fix the coupling constant $\alpha_1$ and increase the decoupling constant $\beta$ or fixing $\beta$ and increasing $\alpha_1$, the mass is decreasing in both scenario. But right panel shows opposite behavior means when $\beta$ increases for fix $\alpha_1$, the mass increases while it is decreases for fixing $\beta$ and increasing $\alpha$. For the solution  \ref{sec3.1},  the maximum mass is achieved at lower values of ($\alpha_1, \beta$) and the region with $\beta<0.024$ is forbidden as the mass becomes imaginary for $\alpha=0.7$. However, the maximum mass for solution  \ref{sec3.2} is obtained at lower $\alpha_1$ and higher $\beta$ value on the $\alpha_1-\beta$ plan.     

Next, Figure \ref{fig9} represents the $\alpha_1-\beta$ plane for equi-$\mathcal{B}_g$ contours. Here we see that the decoupling constant $\beta$ shows very low effect on the bag constant value in solution  \ref{sec3.1}  while totally negligible for solution  \ref{sec3.2}.  But, the bag constant is strongly affected by the constant $\alpha_1$ only for both solutions \ref{sec3.1} (when $\theta_0^0=\rho$) and \ref{sec3.2} (when $\theta_1^1=p_r$). In the first solution, as $\alpha_1$ increases from $0.7$ to $1.2$ the bag constant values lie in the range $60 ~MeV/fm^3$ to $90~MeV/fm^3$ while for second solution, the $\mathcal{B}_g$ lies in a range $50-86~MeV/fm^3$. Hence, we can conclude that for a high value of $\alpha_1$, the bag constant $\mathcal{B}_g$ takes higher value for the solution \ref{sec3.1} than the solution \ref{sec3.2}.

 In order to observe the effects of $\mathcal{B}_g$ and $\alpha_1$ for a fixed radius, Figure \ref{fig10} has been plotted contours of equi-mass in $\mathcal{B}-\alpha_1$ plane for both the solutions when $\theta_0^0$ mimics energy density  (left panel) and when $\theta_1^1$ mimics radial pressure (right panel). We can observe from this Figure \ref{fig10} that the mass of the stellar structure increases in both solutions for increasing $\mathcal{B}_g$ and fixed fixed $\alpha_1$. However,  fixing the bag constant $\mathcal{B}_g$ and increasing the strength of the constant  $\theta_1^1$ , the mass in both solutions decreases. Hence, for obtaining a stellar configuration with high mass  in both solutions the bag constant $\mathcal{B}_g$ must be high and  $\alpha_1$ must be low (see Figure \ref{fig11}).  

  Furthermore, we have also shown the effect of $\mathcal{B}_g$ and $\beta$ on the mass by plotting of Figure \ref{fig11} for contours of equi-mass in $\mathcal{B}_g-\beta$ plane. We see that by fixing the decoupling constant $\beta$ and increasing the $\mathcal{B}_g$ the mass is increasing in both solutions but when fix the bag
 
 constant and $\beta$ move from $0$ to $1$, the mass decreases in first solution \ref{sec3.1} (left panel) while it is increasing for second solution \ref{sec3.2} (right panel). Hence, the heavily massive stellar configuration can be achieved in solution \ref{sec3.1}  for higher $\mathcal{B}_g$ with lower $\beta$ while for second solution \ref{sec3.2} the higher values of $\mathcal{B}_g$ and $\beta$ are needed. It must be noted that for solution 1, there is a forbidden region of where the mass of the configuration becomes imaginary for $\mathcal{B}_g>82 MeV/fm^3$ and $\beta<0.075$.

    \begin{table*}[!htb]
\centering
\caption{The predicted radii of compact stars LMC X-4, PSR J1614+2230, PSR J0740+6620, and GW190814 for the case $\theta^0_0(r)=\rho(r)$ }\label{table2}
 \scalebox{0.85}{\begin{tabular}{| *{11}{c|} }
\hline
\multirow{3}{*}{Objects} & \multirow{3}{*}{$\frac{M}{M_\odot}$}   & \multicolumn{5}{c|}{{Predicted $R$ km}} & \multicolumn{4}{c|}{{Predicted $R$ km}} \\
\cline{3-11}
&& \multicolumn{5}{c|}{$\alpha_1$} & \multicolumn{4}{c|}{$\beta$} \\
\cline{3-11}
&  & 0.8 & 0.9 & 1.0 & 1.1 & 1.2 & 0 & 0.01& 0.02 & 0.03   \\ \hline
LMC X-4 \citep{star1}  &  1.29 $\pm$ 0.05  & $9.86_{-0.034}^{+0.03}$  &  $10.54_{-0.01}^{+0.02}$  &   11.13  &  $11.66_{-0.01}^{+0.01}$  &  $12.14_{-0.01}^{+0.02}$ & $12.04_{-0.01}^{+0.01}$ & $12.09_{-0.01}^{+0.01}$ & $12.14_{-0.01}^{+0.02}$ & $12.20_{-0.01}^{+0.01}$  \\
\hline
 PSR J1614+2230\citep{Ayan2} & 1.97$\pm$0.04  & -  & - & 10.91$_{-0.034}^{+0.04}$ & 11.60$_{-0.01}^{+0.02}$ & 12.18$_{-0.01}^{+0.01}$ & 12.05$_{-0.01}^{+0.01}$ & 12.11 & 12.18$_{-0.01}^{0.01}$ & 12.24$_{-0.01}^{+0.01} $ \\
\hline
PSR J0740+6620 \citep{star3} & $2.14^{+0.2}_{-0.17}$ & - & -  & 10.67$_{-0.0}^{+0.026}$ & 11.54$_{-0.15}^{+0.17}$ & 12.15$_{-0.06}^{+0.0.03}$ & 12.00$_{-0.07}^{+0.0.04}$ & 12.08$_{-0.07}^{+0.04}$ & 12.15$_{-0.06}^{+0.03}$ & 12.22$_{-0.05}^{+0.0.04}$  \\
\hline
GW190814 \citep{wenbin} & 2.5-2.67 & - & -  & - & 10.78$_{-0.0}^{+0.36}$ & 11.97$_{-0.08}^{+0.04}$ & 11.73$_{-0.0}^{+0.06}$ & 11.85$_{-0.11}^{+0.06}$ & 11.96$_{-0.21}^{+0.05}$ & 12.07$_{-0.05}^{+0.04}$  \\
\hline
\end{tabular}}
\end{table*}

\begin{table*}[!htb]
\centering
\caption{The predicted radii of compact stars LMC X-4, PSR J1614+2230, PSR J0740+6620, and GW190814 for the case $\theta^1_1(r)=p_r(r)$ }\label{table3}
 \scalebox{0.83}{\begin{tabular}{| *{12}{c|} }
\hline
\multirow{3}{*}{Objects} & \multirow{3}{*}{$\frac{M}{M_\odot}$}   & \multicolumn{5}{c|}{{Predicted $R$ km}} & \multicolumn{5}{c|}{{Predicted $R$ km}} \\
\cline{3-12}
&& \multicolumn{5}{c|}{$\alpha_1$} & \multicolumn{5}{c|}{$\beta$} \\
\cline{3-12}
&  & 0.8 & 0.9 & 1.0 & 1.1 & 1.2 & 0 & 0.005 & 0.01 & 0.015 & 0.02   \\ \hline
LMC X-4 \citep{star1}  &  1.29 $\pm$ 0.05 & $10.40_{-0.07}^{+0.07}$  &  10.90$_{-0.09}^{+0.09}$  &  11.35$_{-0.09}^{+0.09}$ &   11.79$_{-0.1}^{+0.09}$ & 12.19$_{-0.1}^{+0.08}$ & 11.70$_{-0.1}^{+0.11}$ & 11.79$_{-0.09}^{+0.1}$ & 11.86$_{-009.}^{+0.09}$ & 11.91$_{-0.1}^{+0.12}$ & 12.01$_{-0.04}^{+0.07}$   \\
\hline
 PSR J1614+2230\citep{Ayan2} & 1.97$\pm$0.04  & 11.06$_{-0.02}^{+0.02}$ & 11.68$_{-0.01}^{+0.01}$ & 12.25$_{-0.03}^{+0.02}$ & 12.77$_{-0.03}^{+0.02}$ & 13.26$_{-0.01}^{+0.03}$ & 12.69$_{-0.01}^{+0.03}$ & 12.76$_{-0.01}^{+0.02}$ & 12.82$_{-0.01}^{+0.05}$ & 12.90$_{-0.05}^{+0.11}$ & 12.97$_{-0.02}^{+0.03}$   \\
\hline
PSR J0740+6620 \citep{star3} & $2.14^{+0.2}_{-0.17}$ & 11.07$_{-0.04}^{+0.09}$ &  11.73$_{-0.07}^{+0.01}$ & 12.33$_{-0.12}^{+0.04}$ & 12.89$_{-0.015}^{+0.08}$ & 13.40$_{-0.19}^{+0.11}$ & 12.82$_{-0.17}^{+0.08}$ & 12.90$_{-0.17}^{+0.07}$ & 12.95$_{-0.15}^{+0.07}$ & 13.02$_{-0.15}^{+0.07}$ & 13.09$_{-0.15}^{+0.07}$  \\
\hline
GW190814 \citep{wenbin} &  2.5-2.67 & 10.42$_{-0.0}^{+0.29}$ & 11.53$_{-0.3}^{+0.07}$ & 12.29$_{-0.06}^{+0.05}$ & 12.96$_{-0.04}^{+0.01}$ & 13.55$_{-0.01}^{+0.01}$ & 12.90$_{-0.03}^{+0.02}$ & 12.96$_{-0.02}^{+0.04}$ & 13.0$_{-0.04}^{+0.04}$ & 13.05$_{-0.03}^{+0.06}$ & 13.11$_{-0.07}^{+0.01}$ \\
\hline
\end{tabular}}
\end{table*} 

  \section{Discussion of Findings}\label{sec6}
    In this work we attempted to model the secondary component of the peculiar GW190814 event which to date, has the most unusual mass ratio of the protagonists giving rise to these gravitational signals. Our main interest was to produce a model of the secondary component of the binary merger ie., a neutron star above the accepted mass limit of 2.67 $M_{\odot}$. To this end we modelled compact objects within the framework of $f(\mathcal{Q})$ theory of gravity in addition to exploring the effects of anisotropy introduced via Minimum Gravitational Decoupling method. To close the system of equations we make use of the Tolman IV ansatz for one of the metric functions together with the MIT Bag EoS. We obtained two classes of exact solutions \ref{sec3.1}:  $\theta^0_0=\rho$ - sector and \ref{sec3.2}: $\theta^1_1=p_r$ - sector. These models were subjected to rigorus physical viability tests by independently varying the metricity parameter, $\cal Q$ and the decoupling constant, $\beta$. Our analyses of the density and pressure profiles of our models clearly demonstrated the impact of $\cal Q$ and $\beta$. We found that contributions from the metricity factor, $\cal Q$ increases the densities and stresses within the fluid configurations. The decoupling parameter tended to suppress any variation in density of the compact objects, particularly at the stellar surfaces. The anisotropy parameter for the $\theta^0_0=\rho$ - sector, changed sign as one moves from the center of the star to the boundary. This could characterise a possible phase transition with the change in sign of the anisotropy parameter. For the $\theta^1_1=p_r$ - sector, the anisotropy factor is positive at each interior point of the stellar configuration. This repulsive force due to anisotropy tends to stabilize the object against the inwardly driven gravitational force. We also noted the anisotropy generated in the  $\theta^1_1=p_r$ models is at the maximum $50\%$ greater in magnitude than their $\theta^0_0=\rho$ counterparts. The mass profiles for both classes of solutions are well-behaved and exhibit physically sound behaviour. The novelty of our work can be ascertained in Figures \ref{fig6} and \ref{fig7}. While both classes of solutions account for the masses of well-known compact objects such as LMC X-4, PSR J1614+2230 and PSR J0740+6620, they also predict masses above 2.0 $M_{\odot}$, the observed limiting mass for neutron stars. For the $\theta^0_0=\rho$, predicted masses range from 2.5$M_{\odot}$ to 2.67$M_{\odot}$ with radii less than $12$ km. In the $\theta^1_1=p_r$ models, similar masses occur for radii ranging between 13.11$_{-0.07}^{+0.01}$km and  13.55$_{-0.01}^{+0.01}$km. The contributions from the metricity factor, $\cal Q$ allows for bigger self-gravitating configurations (larger radii) with larger masses. The effect of varying the decoupling constant while holding $\cal Q$ fixed yields slightly smaller radii for the secondary component of the GW190814 event. This allows us to speculate that a combination of contributions from the metricity factor and the decoupling parameter allows for higher mass neutron stars which are stable and may be the secondary progenitor of the black hole-neutron star coalescence which was the birth cry of the GW190814 event.
   
     \section*{Acknowledgement}
The author SKM acknowledges that this work is carried out under TRC Project (Grant No. BFP/RGP/CBS-/19/099), the Sultanate of Oman. SKM is thankful for continuous support and encouragement from the administration of University of Nizwa. 
\begin{widetext}
\section*{Appendix}
\begin{small}
\begin{eqnarray}
&&\hspace{-0.5cm}\theta_{11}(r)=-  B^2 r^8 \alpha_2 + A^2 (2 r^2 \alpha_1 - r^4 \alpha_2) + A [(6 + 4 B r^4) \alpha_1 - 2 r^2 (1 + B r^4) \alpha_2],\nonumber\\
&&\hspace{-0.5cm}\theta_{12}(r)=\big(-4 \mathcal{B}_g \left(r+A r^3+B r^5\right)^2+\left(3+9 A r^2+4 A^2 r^4+11 B r^4+8 A B r^6+4 B^2 r^8\right) \alpha_1 -2 \left(r+A r^3+B r^5\right)^2 \alpha_2 \big),\nonumber\\
&&\hspace{-0.5cm}\theta_{21}(r)=32 \mathcal{B}^2_g r^2 \left(1+A r^2+B r^4\right)^4\,(A r^2+3 B r^4-1)-150 B r^2 \alpha_1 ^2-194 B^2 r^6 \alpha_1 ^2+310 B^3 r^{10} \alpha_1 ^2 +250 B^4 r^{14} \alpha_1 ^2\nonumber\\&&\hspace{0.5cm}+24 B^5 r^{18} \alpha_1 ^2+9 \alpha_1  \alpha_2 +88 B r^4 \alpha_1  \alpha_2 +10 B^2 r^8 \alpha_1  \alpha_2 -232 B^3 r^{12} \alpha_1  \alpha_2 -187 B^4 r^{16} \alpha_1  \alpha_2 -24 B^5 r^{20} \alpha_1  \alpha_2 -2 r^2 \alpha_2 ^2\nonumber\\&&\hspace{0.5cm}-2 B r^6 \alpha_2 ^2+12 B^2 r^{10} \alpha_2 ^2+28 B^3 r^{14} \alpha_2 ^2+22 B^4 r^{18} \alpha_2 ^2+6 B^5 r^{22} \alpha_2 ^2+2 A^5 r^8 \left(-2 \alpha_1 +r^2 \alpha_2 \right)^2+4 \mathcal{B}_g \left(1+A r^2+B r^4\right)^2 \nonumber\\&&\hspace{0.5cm} \times \big[2 \big(9+25 B r^4+A^3 r^6-25 B^2 r^8-9 B^3 r^{12}+A^2 \left(11 r^4-7 B r^8\right)+A r^2 (27-6 B r^4-17 B^2 r^8)\big) \alpha_1 +5 (A r^2+3 B r^4\nonumber\\&&\hspace{0.5cm}-1) \left(r+A r^3+B r^5\right)^2 \alpha_2 \big]+A^4 r^6 \left(\left(46+56 B r^4\right) \alpha_1 ^2-r^2 \left(41+56 B r^4\right) \alpha_1  \alpha_2 +2 r^4 \left(3+7 B r^4\right) \alpha_2 ^2\right)+2 A \big[(-27\nonumber\\&&\hspace{0.5cm}-180 B r^4+197 B^2 r^8+354 B^3 r^{12}+52 B^4 r^{16}) \alpha_1 ^2+r^2 \left(27+55 B r^4-199 B^2 r^8-279 B^3 r^{12}-52 B^4 r^{16}\right) \alpha_1  \alpha_2 \nonumber\\&&\hspace{0.5cm} +r^4 \left(1+B r^4\right)^3 \left(-3+13 B r^4\right) \alpha_2 ^2\big]+2 A^3 r^4 \big[3 \left(1+50 B r^4+24 B^2 r^8\right) \alpha_1 ^2-r^2 \left(11+133 B r^4+72 B^2 r^8\right) \alpha_1  \alpha_2 \nonumber\\&&\hspace{0.5cm} +2 r^4 \left(1+10 B r^4+9 B^2 r^8\right) \alpha_2 ^2\big]+2 A^2 r^2 \big\{\left(-51+53 B r^4+356 B^2 r^8+88 B^3 r^{12}\right) \alpha_1 ^2-2 r^2 (-14+47 B r^4\nonumber\\&&\hspace{0.5cm}+149 B^2 r^8+44 B^3 r^{12}) \alpha_1  \alpha_2 +2 r^4 \left(-1+11 B r^4\right) \left(\alpha_2 +B r^4 \alpha_2 \right)^2\big\}\nonumber\\
&&\hspace{-0.5cm} \theta_{22}(r)=\big\{4 \mathcal{B}_g \left(r+A r^3+B r^5\right)^2-\left(3+9 A r^2+4 A^2 r^4+11 B r^4+8 A B r^6+4 B^2 r^8\right) \alpha_1 +2 \left(r+A r^3+B r^5\right)^2 \alpha_2 \big\}^2,\nonumber\\
&&\hspace{-0.5cm} \theta_{23}(r)=-2 B^2 r^6 \alpha_1 +\alpha_2 +2 B r^4 \alpha_2 +B^2 r^8 \alpha_2 +A^2 \left(-2 r^2 \alpha_1 +r^4 \alpha_2 \right)+2 A \left(-3 \alpha_1 -2 B r^4 \alpha_1 +r^2 \alpha_2 +B r^6 \alpha_2 \right), \nonumber\\
&&\hspace{-0.5cm}\mathcal{B}_{g1} (R) =\frac{\beta \left(4 A^2 R^4+A R^2 \left(8 B R^4+9\right)+4 B^2 R^8+11 B R^4+3\right) \left(-6 \alpha_1  A+A \alpha_2  R^2+\alpha_2 +\alpha_2  B R^4 -6 \alpha_1  B R^2\right)}{\left(A R^2+B R^4+1\right)^2}\nonumber\\ &&\hspace{0.5cm}-\frac{6 \alpha_1  \left(R^2 \left(A^2+5 B\right)+2 A B R^4+3 A+B^2 R^6\right)}{\left(A R^2+B R^4+1\right)^2}-\frac{2 \alpha_2  \beta R^2 \left(-6 \alpha_1  A+A \alpha_2  R^2+\beta +\beta  B R^4-6 \alpha_1  B R^2\right)}{\alpha_1 },\nonumber\\
&&\hspace{-0.5cm} \mathcal{B}_{g2} (R)=\left(A \beta R^2+B \beta R^4+1\right),~~~~\mathcal{M}_1(R)=\left(A+B R^2\right) \left(-\Lambda R^2+3 \beta+3\right)-\Lambda, \nonumber\\
&&\hspace{-0.5cm}\mathcal{C}_{11}(R)=R^2\,\left(-12 \alpha_1 A+2 \mathcal{B}_g \left(3 A R^2+2 B R^4+6\right)+3 A \alpha_2  R^2+6 \alpha_2 +2 \alpha_2  B R^4-6 \alpha_1 B R^2\right)+18 \alpha_1 \ln \mathcal{C}_{12}(R),\nonumber
\end{eqnarray}
\begin{eqnarray}
&&\hspace{-0.5cm} \mathcal{C}_{12}(R) = \frac{\alpha_2  \beta R^2 \left(A R^2+B R^4+1\right)-6 \alpha_1 \left(A \beta R^2+B \beta R^4-1\right)}{6 \alpha_1 \left(A R^2+B R^4+1\right)},~~\mathcal{M}_{21}(R)=-[4 \mathcal{B}_g +2 \alpha_2] \left(A R^3+B R^5+R\right)^2,\nonumber\\
&&\hspace{-0.5cm} \mathcal{M}_{2}(R)=\frac{\beta\, R^2 \left(-2 \alpha_1  \left(R^2 \left(A^2+5 B\right)+2 A B R^4+3 A+B^2 R^6\right)+8 \mathcal{B}_g \left(A R^2+B R^4+1\right)^2+\alpha_2 \left(A R^2+B R^4+1\right)^2\right)}{2 \left(A R^2+B R^4+1\right) \left[\alpha_1  \left(R^4 \left(4 A^2+11 B\right)+8 A B R^6+9 A R^2+4 B^2 R^8+3\right)+\mathcal{M}_{21}(R)\right]},\nonumber\\
&&\hspace{-0.5cm} C_{21}(R)=\frac{\beta\, R^2 \left(-2 \alpha_1  \left(R^2 \left(A^2+5 B\right)+2 A B R^4+3 A+B^2 R^6\right)+8 \mathcal{B}_g \left(A R^2+B R^4+1\right)^2+\alpha_2 \left(A R^2+B R^4+1\right)^2\right)}{\alpha_1  \left(4 A^2 R^4+8 A B R^6+9 A R^2+4 B^2 R^8+11 B R^4+3\right)-4 \mathcal{B}_g \left(A R^3+B R^5+R\right)^2-2 \alpha_2 \left(A R^3+B R^5+R\right)^2},\nonumber
\end{eqnarray}
\end{small}
\end{widetext}

\end{document}